\documentclass[journal]{new-aiaa}
\usepackage[utf8]{inputenc}
\usepackage{textcomp}

\usepackage{graphicx}
\usepackage{pgfplots}
\usepackage{tikz}
\usetikzlibrary{matrix}
\usepackage{amsmath}
\usepackage[version=4]{mhchem}
\usepackage{siunitx}
\usepackage{longtable,tabularx}
\usepackage{bm}
\usepackage{diagbox}
\newcommand{\bA}[1]{\mathcal{A}^{(#1)}}
\newcommand{\bH}[1]{\mathcal{H}^{(#1)}}
\newcommand{\bY}[1]{\mathcal{Y}^{(#1)}}

\newcommand{\ba}[1]{\bm{a}^{(#1)}}

\newcommand{\bd}[1]{\bm{d}^{(#1)}}

\newcommand{\bc}{\bm{c}}

\newcommand{\bu}{\bm{u}}
\newcommand{\bv}{\bm{v}}
\newcommand{\bx}{\bm{x}}

\newcommand{\bxi}{\bm{\xi}}
\newcommand{\bdelta}{\bm{\delta}}

\newcommand{\pp}[2]{\frac{\partial #1}{\partial #2}}
\newcommand{\sym}{\mbox{sym}}
\newcommand{\tr}{\mbox{tr}}

\newcommand{\ma}{\mathit{Ma}}
\newcommand{\pr}{\mathit{Pr}}
\newcommand{\kn}{\mathit{Kn}}

\title{Lattice Boltzmann simulation of non-equilibrium flows using spectral
	multiple-relaxation-time collision model}

\author{Su Yan\footnote{Ph.~D. Student, Department of Mechanics and Aerospace
		Engineering.}}
\affil{Southern University of Science and Technology, 518055 Shenzhen, People’s
	Republic of China}

\author{Xiaowen Shan\footnote{Chair Professor, IAS}}
\affil{BNU-HKBU United International College, 519087 Zhuhai, People’s
	Republic of China}

\begin{document}

\maketitle

\begin{abstract}
Prediction of non-equilibrium flows is critical to space flight. In the present
work we demonstrate that the recently developed spectral
multiple-relaxation-time (SMRT) lattice Boltzmann (LB) model is theoretically
equivalent to Grad's eigen-system [Grad, H., In Thermodynamics of Gases (1958)]
where the eigen-functions obtained by tensor decomposition of the Hermite
polynomials are also those of the linearized Boltzmann equation.  Numerical
results of shock structure simulation using Maxwell molecular model agree very
well with those of a high-resolution fast spectral method code up to Mach 7
provided that the relaxation times of the irreducible tensor components match
their theoretical values.  If a reduced set of relaxation times are used such as
in the Shakhov model and lumped-sum relaxation of Hermite modes, non-negligible
discrepancies starts to occur as Mach number is raised, indicating the necessity
of the fine-grained relaxation model.  Together with the proven advantages of
LB, the LB-SMRT scheme offers a competitive alternative for non-equilibrium flow
simulation.
\end{abstract}

\section*{Nomenclature}

\noindent(Nomenclature entries should have the units identified)

{\renewcommand\arraystretch{1.0}
\noindent
\begin{longtable*}{@{}l @{\quad=\quad} l@{}}
	$\ba{n}$         & Hermite expansion coefficient of distribution                \\
	$B$              & Collision kernel                                             \\
	$\bm{c}$         & peculiar velocity, $\bm{\xi}-\bm{u}$                         \\
	$\bd{n}$         & local Hermite expansion coefficient of distribution          \\
	$f$              & velocity distribution function                               \\
	$f_{eq}$         & Maxwell-Boltzmann distribution                               \\
	$f_{ES}$         & Gaussian distribution of ES-BGK model                        \\
	$f_{S}$          & target distribution of Shakhov model                         \\
	$g$              & scaled distribution, $g=\theta^{3/2}f/\rho$                  \\
	$\bm{k}$         & shifted constant integer vector                              \\
	$\kn$            & Knudsen number                                               \\
	$L$              & linearized Boltzmann collision operator                      \\
	$L_{GJ}$         & Gross-Jackson model                                          \\
	$L_{n}^{a}$      & Associated Laguerre polynomial                               \\
	$\bm{m}$         & 3rd-order rank-3 traceless moment, $\int f\bH{3,0}(\bc)d\bc$ \\
	$\ma$            & Mach number                                                  \\
	$p$              & Hydro-static pressure                                        \\
	$\bm{P}$         & pressure tensor, $\int f\bm{c}^2 d\bm{c}$                    \\
	$\bm{P}^{(n,m)}$ & spherical moment                                             \\
	$P_l$            & Legendre polynomial                                          \\
	$\pr$            & Prandtl number                                               \\
	$\bm{q}$         & heat flux                                                    \\
	$\bm{Q}$         & 3rd-order moment, $\int f\bm{c}^3 d\bm{c}$                   \\
	$\bm{R}$         & 4th-order moment, $\int f\bm{c}^4 d\bm{c}$                   \\
	$S_{l+1/2}^n$    & Sonine polynomial                                            \\
	$t$              & time, $s$                                                    \\
	$\bm{u}$         & macroscopic velocity                                         \\
	$\bm{v}$         & temperature scaled peculiar velocity, $\bc/\sqrt{\theta}$    \\
	$w$              & weight of discrete velocity                                  \\
	$\bx$            & space position, $m$                                          \\
	$Y^{m}_l$        & Spherical harmonic                                           \\
	$\gamma$         & Specific heat ratio                                          \\
	$\bdelta$        & Kronecker delta                                              \\
	$\epsilon$       & Azimuthal angle                                              \\
	$\theta$         & temperature                                                  \\
	$\lambda_{nl}$   & Eigenvalue of linearized BCO                                 \\
	$\lambda_{nm}$   & Eigenvalue of Grad's linearized BCO                          \\
	$\nu$            & Kinematic viscosity                                          \\
	$\bxi$           & molecular velocity                                           \\
	$\bxi_r$         & relative velocity                                            \\
	$\rho$           & density                                                      \\
	$\bm{\sigma}$    & Stress tensor                                                \\
	$\tau$           & Relaxation time                                              \\
	$\chi$           & Polar angle                                                  \\
	$\psi$           & Perturbed distribution                                       \\
	$\bm{\Psi}$      & Eigen-function of Linearized BCO                             \\
	$\Omega$         & collision operator                                           \\
	$\omega$         & weight function of Hermite polynomials                       \\
	$\bH{n}$         & Hermite polynomial                                           \\
	$\bY{n}$         & Solid homogeneous spherical harmonic in Cartesian coordinate \\
	\multicolumn{2}{@{}l}{Subscripts}                                               \\
	$'$              & post-collision values
\end{longtable*}}

\section{Introduction}

\lettrine{N}{on-equilibrium} fluid flows are ubiquitous in nature and
engineering applications where the constituent molecules fail to reach thermal
equilibrium on the temporal or spatial scale of interest due to the relatively
large collision interval or
mean-free-path~\cite{Ivanov1998,Gravesen1993,Javadpour2021}.  As the
Naiver-Stokes-Fourier (NSF) equations break down in such limits, the Boltzmann
equation is commonly used to determine the fluid's behavior.  The high
dimensionality of the distribution function and the complexity of the Boltzmann
collision operator (BCO) make the computational solution of the Boltzmann
equation a long-standing challenge in both applied mathematics and computational
physics.

The most popular solution method is perhaps the direct simulation Monte Carlo
(DSMC)~\cite{Bird1994} method which uses a molecular dynamic approach in
simulating the behavior of the Boltzmann equation. DSMC requires the cell size
and time step to be much smaller than the mean free path and mean collision
interval and poses significant limitations on its use in continuous
regime~\cite{Baker2005}.  To reduce of the high-dimensional distribution
function, many deterministic numerical methods~\cite{Dimarco2014} have been
proposed.  The discrete velocity method
(DVM)~\cite{goldstein1989investigations,bobylev1995approximation} uses a fixed
set of discrete velocities to approximate the velocity space and construct
discrete collision mechanism on the velocity nodes to preserve the main physical
properties of BCO.  However, the accuracy is limited and computational cost high
when applied to the full BCO~\cite{Dimarco2014}.  The spectral
method~\cite{Bobylev1988,Pareschi1996,Pareschi2000} expands the distribution by
Fourier series in velocity space to achieve spectral accuracy. The fast spectral
method (FSM)~\cite{Mouhot2006} further reduces the computational cost and was
successfully applied to non-equilibrium flows~\cite{Wu2013,Wu2014}.  In addition
to Fourier expansion, Hermite and Burnett polynomials have recently been used to
approximate the distributions~\cite{Wang2019,Cai2020}.  Instead of solving for
the distribution function itself, the moment method~\cite{Grad1949a} solves for
the low-order moments of the distribution that dictates the macroscopic
hydrodynamics.  The Hermite expansion of the distribution is substituted into
the Boltzmann equation and closed at various orders and
closures~\cite{Struchtrup2003b}.

To simplify the BCO, many models have been suggested among which the simplest is
the Bhatnagar–Gross–Krook (BGK) model~\cite{Bhatnagar1954} in which the
collision is modeled as a uniform relaxation of the distribution towards its
equilibrium.  The ellipsoidal statistical BGK (ES-BGK)~\cite{Holway1966} and the
Shakhov model~\cite{Shakhov1968} modifies the target of relaxation to restore
the correct Prandtl number. These kinetic models perform well in near continuous
regime but become less accurate in transition regime~\cite{Fei2020}. To improve
the simulation accuracy of transition flows, an extended Shakhov model was
proposed with incorporating more relaxation of higher-order
moments~\cite{Yao2023}. However, how to consider the relaxation process
of higher-order moments is still an issue worth exploring.

In the past three decades the lattice Boltzmann method
(LBM)~\cite{Chen1998a,Benzi1992}, originated from a simple cellular automaton
fluid model~\cite{Frisch1986}, has been developed into a practical computational
fluid dynamics (CFD) tool mostly independent of the classic kinetic theory. Thus
far its development focuses on the recovery of the Navier-Stokes hydrodynamics
using the BGK collision model~\cite{Chen1992,Qian1992}, first for
near-incompressible flows and more recently for thermo-compressible flows. After
a formal connection with classical kinetic theory was
established~\cite{Shan1998,Shan2006b}, the LBM can be recognized as a special
DVM scheme obtained by first projecting the Boltzmann-BGK equation into the
Hilbert space spanned by the leading Hermite polynomials and then evaluating at
the abscissas of a Gauss-Hermite quadrature so that the kinetic moments of the
truncated distribution are precisely represented by the discrete distribution
function.  On the other hand the LBM can also be viewed as a computation scheme
of the moment method.  This connection has paved the way for simulating
non-equilibrium flows by allowing more elaborated collision models to be
developed so that the Boltzmann equation itself, rather than its hydrodynamics,
can be simulated by LBM.  In recent years, a spectral-space
multiple-relaxation-time (SMRT) collision model was developed where the
distribution is expanded into a finite Hermite series in local frame. Each of
the Hermite polynomials is further decomposed into components corresponding to
an irreducible representation of SO(3)~\cite{Zee2016} which are the minimum
groups closed under spatial rotation.  The irreducible tensors are then relaxed
with independent relaxation times without violating rotation
symmetry~\cite{Shan2007,Shan2019,Li2019,Li2020b,Shi2021,Shan2021}.

A common benchmark problem for non-equilibrium flows is the computation of shock
structure.  NSF equations only give acceptable results for very small Mach
number~\cite{Gilbarg1953,Pham-Van-Diep1991}.  Higher order hydrodynamic models,
\textit{e.g.}, Burnett-type equations yield better results~\cite{Foch1973} but
encounter instability and thermodynamic
inconsistency~\cite{bobylev1982chapman,Comeaux1995}.  For moment method, Grad
first used the 13-moment equations to obtain reasonable results for low Mach
numbers but artifact known as subshock occurs if $\ma > 1.65$~\cite{Grad1952}. 
Hydrodynamic equations based on Ikenberry's spherical moments and thrid-order
Maxwell iteration~\cite{Ikenberry1956,Truesdell1980a} were used to solve the
shock structure problem with no superiority over NSF
observed~\cite{street1960shock}. By combining the moment method and
Chapman-Enskog expansion~\cite{Chapman1954}, the regularized moment
method~\cite{Struchtrup2003b} can provide acceptable results for $\ma<3.0$ and
always continuous smooth results over a wide range of Mach
number~\cite{Torrilhon2004,Timokhin2017}, but satisfactory quantitative results
cannot be obtained for high Mach number. If the accuracy of high $\ma$ shocks is
required, the number of moments must be increased or nonlinear closure must be
used~\cite{Torrilhon2016}. For methods such as DMSC and DVM that directly solve
the Boltzmann equation, accurate solutions of shock wave structures can be
obtained.

The present work has two purposes.  First we show that the SMRT model is a
discrete incarnation of the Grad's eigen-system which, as far as we know, has
not been used as the basis of any computation.  This correspondence not only
presents a theoretical framework within which the SMRT model can be scrutinized
for further development, but also provides a comprehensive numerical tool for
moments equations that enables simulation of three-dimensional unsteady
non-equilibrium flows in complex geometry.  Secondly we numerically validate the
LB-SMRT scheme for computation of highly non-equilibrium flows using the shock
structure problem as a benchmark.  We show that by properly choosing the
reference temperature so that the Holway's criterion~\cite{Holway1964} is
satisfied, and by using the quadrature rules in a moving reference
frame~\cite{Frapolli2016,Shi2021phd}, shock structure can be computed to $\ma=7$
while the results remain in excellent agreement with that of a fast spectral
method computation.

The rest of the paper is organized as follows.  In Sec.~\ref{sec:kinetic} we
briefly review the simplified collision models to which the LB-SMRT model is
related in one way or another.  In Sec.~\ref{sec:smrt} the LB-SMRT model is
reviewed with respect to the kinetic models.  In Sec.~\ref{sec:numerical} we
present simulation results of shock structure using LB-SMRT.  In particular,
dependence on truncation order, comparison with the Shakhov model, and the
necessity of independent relaxation of the irreducible components are discussed.
Finally in Sec.~\ref{sec:conclusion} some conclusions and future prospects are
given.

\section{Collision models in kinetic theory}

\label{sec:kinetic}

The single particle velocity distribution function, $f(\bx,\bxi,t)$, is defined
in the phase-space $(\bx, \bxi)$, where $\bx$ and $\bxi$ are the space and
velocity coordinates respectively, such that $f(\bx, \bxi, t)d\bxi d\bx$ is the
probability of finding the particle in the element $d\bxi d\bx$ at time $t$.
Omitting external body force, the Boltzmann equation describing the evolution of
$f$ can be written as:
\begin{equation}
    \pp ft + \bxi\cdot\nabla f = \Omega(f)
    \label{eq:BE}
\end{equation}
where $\Omega(f)$ is the Boltzmann collision operator (BCO) which is a complex
integral-differential operator:
\begin{equation}
    \label{eq:BCO}
    \Omega(f)=\int B(\chi,|\bm{\xi}_r|)
    \left[f(\bxi')f(\bxi'_\ast)-f(\bxi)f(\bxi_\ast)\right]d\chi d\epsilon d\bxi_\ast
\end{equation}
where $\chi$ and $\epsilon$ are respectively polar and azimuthal angles defining
the unit vector with respect to polar direction. $\bm{\xi}_r$ is the relative
velocity of collision molecules. $B(\chi,|\bm{\xi}_r|)$ is the collision kernel
related to inter-molecular potentials. $\bm{\xi}'$ and $\bm{\xi}'_{*}$ are
corresponding post-collision velocities.  At equilibrium, the distribution takes
the form of the Maxwell-Boltzmann equilibrium distribution:
\begin{equation}
	\label{eq:Maxwellian}
	f^{eq}(\rho,\bu,\theta) = \frac{\rho}{(2\pi\theta)^{3/2}}\exp
	\left[-\frac{|\bxi-\bu|^2}{2\theta}\right]
\end{equation}
where $\rho$, $\bu$ and $\theta$ are respectively the dimensionless macroscopic
density, velocity, and temperature in ``energy units,'' \textit{i.e.}, the
temperature is scaled by a constant reference temperature, $T_0$, and all
velocities scaled by the corresponding thermal speed $\sqrt{k_BT_0}$ where $k_B$
is the Boltzmann constant~\cite{Shan2006b}.

\subsection{Linearized Boltzmann equation}

In the development of simplified kinetic models, the linearized Boltzmann
equation plays a significant role.  Assuming the distribution is sufficiently
close to its equilibrium, it can be approximated as
\begin{equation}
	\label{eq:ab Linear VDF}
	f(\bx,\bxi,t) = f^{eq}\left[1+\psi(\bx,\bxi,t)\right]
\end{equation}
where $\psi$ is the perturbation satisfying $|\psi|\ll 1$, and $\rho$, $\bu$ and
$\theta$ in $f_{eq}$ are assumed homogeneous and constant.  On substituting into
Eq.~\eqref{eq:BE} and neglecting terms quadratic in $\psi$, we have the
linearized Boltzmann equation
\begin{equation}
   \frac{\partial \psi}{\partial t}+\bm{\xi}\cdot \nabla \psi=L(\psi)
\end{equation}
with the linearized BCO:
\begin{equation}
	\label{eq:linear-BCO}
	L(\psi)=\int B(\chi, |\bxi_r|)
	\left[\psi(\bxi')+\psi(\bxi'_\ast)-\psi(\bxi)-\psi(\bxi_\ast)\right]f^{eq}
	d\chi d\epsilon d\bxi_\ast
\end{equation}
Furthermore, for Maxwell molecules which repel with force $\sim r^{-5}$, the
collision kernel becomes independent of $|\bxi_r|$ and takes the form of an
eigen system $L(\Psi_{nlm}) = -\lambda_{nl}\Psi_{nlm}$ where $\lambda_{nl}$ is
the eigenvalue~\cite{WangChang1952}.  The eigen-function was obtained in
spherical coordinates $(\xi, \vartheta, \varphi)$ as:
\begin{equation}
	\label{eq:WCU}
	\Psi_{nlm}\sim S^{(n)}_{l+1/2}(\xi)\xi^l Y_l^m(\vartheta,\varphi)
\end{equation}
which is commonly known as the Burnett polynomial~\cite{Burnett1936}.
$S_{l+1/2}^{n}(\xi)$ is known as the Sonine polynomials which, up to a constant
factor, is the associated Laguerre polynomials:
\begin{equation}
	L^a_n(z) = \sum_{k=0}^{n} \frac{(-1)^k(a+n)!}{k!(n-k)!(a+k)!}z^k
\end{equation}
and $Y_l^m(\vartheta, \varphi)$ the \textit{spherical harmonics}.  We note that
$\xi^lY_l^m(\vartheta, \varphi)$ is the \textit{solid spherical harmonics}, the
solution of the Laplace equation.  The first few eigen-functions are related to
the perturbed density, momentum and energy, with the corresponding eigenvalues
$\lambda_{00} = \lambda_{01} = \lambda_{10}=0$ due to the conservation laws.
Additionally, $\lambda_{02}$ and $\lambda_{11}$ are related to the perturbed
stress and heat flux respectively. Based on this eigen-system, Gross and
Jackson~\cite{Gross1959} proposed a model for the linearized BCO which gives the
relaxation rate of any distribution, $\psi$, as:
\begin{align}
	L_{GJ}(\psi) = \lambda_{st}\psi + f^{eq}\sum_{nl}(\lambda_{nl}-\lambda_{st})
	\frac{2l+1}{4\pi}\int P_l(\cos\phi')g_{nl}(\xi)g_{nl}(\xi_\ast)
	\psi(\xi_\ast)d\bxi_\ast
\end{align}
where $P_l$ is the Legendre polynomial with $\phi'$ being the angle between
$\bm{\xi}$ and $\bm{\xi}'$, $g_{nl}\equiv S^{(n)}_{l+1/2}(\xi)\xi^l$ the radial
part of the eigen-function, and $\lambda_{st}$ an arbitrary negative number. For
reasonable choosing of $\lambda_{st}$, the linearized BGK, ES-BGK and Shakhov
models can be recovered~\cite{leiwubook}.

\subsection{Grad's Hermite eigen-system}

Hereinafter the standard notation of symmetric tensor is used.  Defining the
\textit{symmetrization} operator:
\begin{equation}
	\sym(\mathcal{A})\equiv\frac{1}{r!}\sum A_{i_1\cdots i_r}
\end{equation}
where $\mathcal{A}$ is a rank-$r$ tensor and the summation is over the $r!$
permutations of the $r$ indexes, the \textit{symmetric product} of two tensors
$\mathcal{A}$ and $\mathcal{B}$, denoted by $\mathcal{AB}$, is defined as:
\begin{equation}
	\mathcal{AB}=\sym\left(\mathcal{A}\otimes\mathcal{B}\right)
	\label{eq:symm}
\end{equation}
where $\otimes$ stands for the normal tensor product. The $n$-th power of
$\mathcal{A}$ is:
\begin{equation}
	\mathcal{A}^n = \underbrace{\mathcal{A}\cdots\mathcal{A}}_{\text{$n$-fold}}
\end{equation}

Grad~\cite{Grad1949a} proposed to expand the dimensionless distribution in the
local reference frame in terms of Hermite orthogonal polynomials~\cite{Grad1949}
\begin{equation}
	\label{eq:hermite}
	\frac{\theta^{3/2}}{\rho}f(\bx, \bxi, t) \equiv g =\omega(\bv)
	\sum_{n=0}^\infty\frac{1}{n!}\ba{n}(\bx,t):\bH{n}(\bv)
\end{equation}
where $\bv\equiv(\bxi-\bu) / \sqrt{\theta}$, $\bH{n}(\bv)$ is the $n$-th Hermite
polynomial, $\omega(\bv)\equiv (2\pi)^{-3/2} \exp\left(-v^2/2\right)$ the weight
function \textit{w.r.t.}\ which $\bH{n}(\bv)$ are orthogonal to each other, and
`:' denotes full contraction between two tensors.  The expansion coefficients
given by
\begin{equation}
	\label{eq:a}
	\ba{n}(\bx,t) = \int f\bH{n}(\bv) d\bv
\end{equation}
correspond to the macroscopic hydrodynamic variables~\cite{Grad1949a}. The first
few of them are $\ba{0} = 1$, $\ba{1} = 0$, $\ba{2} = \bm{P} / p-\bdelta$,
$\ba{3} = \bm{Q}/p\sqrt{\theta}$, and $\ba{4} = \bm{R}/p\theta - 6\bm{P}\bdelta
/ p + 3\bdelta^2$, where
\begin{equation}
	\bm{P}\equiv\int f\bc^2 d\bc,\quad
	\bm{Q}\equiv\int f\bc^3 d\bc,\quad\mbox{and}\quad
	\bm{R}\equiv\int f\bc^4 d\bc
\end{equation}
are all tensorial velocity moments, where $\bc=\bxi-\bu$ is the
\textit{peculiar} velocity. Particularly, $\bm{P}$ is the pressure tensor, $p =
P_{ii}/3$ the hydro-static pressure, and $q_{i}=Q_{ikk}/2$ the heat flux.
Noticing that $\rho\theta^{-3/2}\omega = f^{eq}$, Eq.~\eqref{eq:hermite} can be
recognized as an expansion about the \textit{local} Maxwellian:
\begin{equation}
	f = f^{eq}\left[1 + \frac 1{2!}\ba{2}:\bH{2} +
	\frac 1{3!}\ba{3}:\bH{3} + \cdots\right]
\end{equation}

On substituting Eq.~\eqref{eq:hermite} into Eq.~\eqref{eq:BCO}, Grad obtained
the Hermite expansion of the BCO.  The $n$-th expansion coefficient is an
infinite quadratic sum of $\ba{n}$:
\begin{equation}
	\label{eq:Grad's expasion of BCI}
	\ba{n}_\Omega = \sum_{r,s=0}^{\infty}\bm{\beta}^{nrs}:\ba{r}\ba{s}
\end{equation}
where $\bm{\beta}^{nrs}$ is a tensor of rank $n+r+s$.  Obviously if $f$ is
approximated by its first $N$ Hermite polynomials, \textit{i.e.}, $\ba{n} = 0$
for $n> N$, the quadratic sum is finite.  Moreover, for Maxwell molecules,
$\bm{\beta}^{nrs} = 0$ for $n\neq r+s$.  The quadratic sum is also finite in
this case, and the Hermite polynomials are the eigen-functions of the linearized
BCO~\cite{Grad1958}.  Particularly for the third-order ($N = 3$) approximation,
Grad obtained a closed system where the state variables are $\ba{0}$, $\ba{1}$,
$\ba{2}$ and part of $\ba{3}$, which is widely known as the 13-moment system.

Alternative to the expansion about the local Maxwellian, the distribution can
also be expanded about the absolute Maxwellian as~\cite{Grad1958}:
\begin{equation}
	\label{eq:hermite1}
	f(\bx, \bxi, t) = \omega(\bxi)
	\sum_{n=0}^\infty\frac{1}{n!}\ba{n}(\bx,t):\bH{n}(\bxi),
\end{equation}
where $\omega(\bxi)$ is the Maxwellian with homogeneous and constant parameters
$\bu = 0$ and $\theta = 1$ independent of any macroscopic properties.  The
leading few expansion coefficients are:
\begin{equation}
	\ba{0} = \rho,\quad
	\ba{1} = \rho\bu,\quad
	\ba{2} = \bm{P} + \rho\bu^2-\bdelta
\end{equation}
On substituting into Eq.~\eqref{eq:linear-BCO}, Grad~\cite{Grad1958} established
that for Maxwell molecules, given an arbitrary trace-less symmetric tensor
$\mathbf{a}$, $\mathbf{a}\bdelta^m$ is an eigen-function of the linearized
BCO. The complete set of eigen-functions can be obtained by a
``\textit{peeling-off}'' procedure which recursively subtracts trace-less
components from $\bH{n}$ and effectively decomposes it into a sum of the
eigen-functions as
\begin{equation}
	\bH{n}(\bxi) = \sum_{m=0}^{\lfloor n/2\rfloor}\bH{n,m}(\bxi)\bdelta^m
\end{equation}
where $\bH{n,m}$ are trace-less symmetric tensors of rank $n-2m$, and the
product is a symmetric product.  As $\bH{n,m}$ are also orthogonal to each other
\textit{w.r.t.}\ $\omega(\bxi)$, Eq.~\eqref{eq:hermite1} can be written as
\begin{equation}
	\label{eq:tensor decompo}
	f(\bx, \bxi, t) = \omega(\bxi)\sum_{n=0}^\infty\frac{1}{n!}
	\sum_{m=0}^{\lfloor n/2\rfloor}\ba{n,m}(\bx,t):\bH{n,m}(\bxi)
\end{equation}

Taking $n=2$ as an example, we have $\bH{2} = \bxi^2-\bdelta$ and $\tr\bH{2} =
\xi^2-3$ where $\tr\mathcal{A}$ stands for the trace of tensor $\mathcal{A}$. We
can hence write $\bH{2} = \bH{2,0} + \bH{2,1}\bdelta$ with
\begin{equation}
	\bH{2,0} = \bH{2} - \frac 13\tr\left(\bH{2}\right)\bdelta
			 = \bxi^2 - \frac{\xi^2}3\bdelta,\quad\mbox{and}\quad
	\bH{2,1} = \frac{\xi^2}{3}-1
\end{equation}
Using the orthogonality relation, we have $\ba{2,0} = \bm{P} - p\bdelta =
\bm{\sigma}$, where $\bm{\sigma}$ is the trace-less deviatoric stress tensor and
$\ba{2,1} = p-\rho$. Results for $n=3$ and $4$ are listed in
Tables~\ref{tab:poly} and \ref{tab:coefficient}.  We note that although
Eq.~\eqref{eq:tensor decompo} is valid in general, for the purposes of
demonstrating linearized BCO, terms related to $\bu$ and higher than first order
in linearization are omitted in Table~\ref{tab:coefficient}.

\begin{table}[hbt!]
\caption{Trace-less components of leading Hermite polynomials}
\label{tab:poly}
\centering
\begin{tabular}{l|l|l|l|l}
	\hline
	$n$ & $\bH{n}(\bxi)$                     & $\bH{n,0}(\bxi)$                                                  & $\bH{n,1}(\bxi)$                                            & $\bH{n,2}(\bxi)$         \\ \hline
	2   & $\bxi^2-\bdelta$                   & $\bxi^2 - \frac 13\xi^2\bdelta$                                   & $\frac 13(\xi^2-3)$                                         &                          \\
	3   & $\bxi^3-3\bxi\bdelta$              & $\bxi^3-\frac{3}{5}\xi^2\bxi\bdelta$                              & $\frac{3}{5}(\xi^2-5)\bxi$                                  &                          \\
	4   & $\bxi^4-6\bxi^2\bdelta+3\bdelta^2$ & $\bxi^4-\frac{6}{7}\xi^2\bxi^2\bdelta+\frac{3}{35}\xi^4\bdelta^2$ & $\frac 67(\xi^2-7)\left(\bxi^2-\frac 13\xi^2\bdelta\right)$ & $\frac 15\xi^4-2\xi^2+3$ \\ \hline
\end{tabular}
\end{table}

\begin{table}[hbt!]
\caption{Trace-less components of leading Hermite expansion coefficients}
\label{tab:coefficient}
\centering
\begin{tabular}{l|c|c|c|c}
	\hline
	$n$ &                $\ba{n}$                &                               $\ba{n,0}$                                &                               $\ba{n,1}$                               &             $\ba{n,2}$             \\ \hline
	2   &                $\bm{P}$                &                              $\bm{\sigma}$                              &                                $p-\rho$                                &                                    \\
	3   &                $\bm{Q}$                &                    $\bm{Q}-\frac{6}{5}\bm{q}\bdelta$                    &                          $\frac{6}{5}\bm{q}$                           &                                    \\
	4   & $\bm{R}-6\bm{P}\bdelta+3\rho\bdelta^2$ & $\bm{R}-\frac{6}{7}tr(\bm{R})\bdelta+\frac{3}{35}tr^2(\bm{R})\bdelta^2$ & $\frac{2}{7}\left[3tr(\bm{R})-tr^2(\bm{R})\bdelta\right]-6\bm{\sigma}$ & $\frac{1}{5}tr^2(\bm{R})-6p+3\rho$ \\ \hline
\end{tabular}
\end{table}

Let $\psi$ in Eq.~\eqref{eq:ab Linear VDF} be expanded like Eq.~\eqref{eq:tensor
	decompo}, the linear BCO can then be written as
\begin{equation}
	\label{eq:Grad's eigen-system}
	L(\psi) = \sum_{n=0}^\infty\frac{1}{n!}
	\lambda_{nm}\sum_{m=0}^{\lfloor n/2\rfloor}\ba{n,m}(\bx,t):\bH{n,m}(\bxi)
\end{equation}
where $\lambda_{nm}$ are the eigenvalues of the eigen-function $\psi_{nm}\equiv
\ba{n,m}:\bH{n,m}$.  It is evident from Table~\ref{tab:coefficient} that
$\lambda_{00} = \lambda_{10} = 0$ due to conservation of mass and momentum;
$\lambda_{20}$ and $\lambda_{21}$ are related respectively to the shear and bulk
viscosity, the latter of which vanishes in a monatomic gas; and $\lambda_{31}$
corresponds to the thermal diffusivity.

The trace-less symmetric tensors, $\bH{n,m}(\bxi)$, have been further developed
by a number of authors. Ikenberry~\cite{Ikenberry1962} gave its explicit
expression as
\begin{equation}
	\bH{n,m}(\bxi) = \frac{(-1)^m n!(2n-4m+1)!!}{(n-2m)!(2n-2m+1)!!}
	L_m^{n-2m+\frac 12}\left(\frac{\xi^2}2\right)\bY{n-2m}(\bxi)
\end{equation}
where $\bY{k}(\bxi)$ is the $k$-th solid spherical harmonics expressed in
Cartesian coordinates~\cite{Ikenberry1961}.  Noticing Eq.~\eqref{eq:WCU}, this
confirms that $\bH{n,m}$ are the eigen-functions of the linearized BCO for
Maxwell molecules. Moreover, Ikenberry~\cite{Ikenberry1962} pointed out that the
$(n+1)(n+2)/2$ distinct components of $\bH{n}$, which provide a
\textit{reducible} representation of SO(3), can be decomposed into $\lfloor
n/2\rfloor+1$ groups, corresponding to the $2n-4m+1$ linearly independent
components of $\bH{n,m}$, $m = 0, \cdots, \lfloor n/2\rfloor$ which furnish an
\textit{irreducible} representation of SO(3), meaning that these groups of
components are the minimum sets closed under spatial rotation.  A method to
decompose a generic symmetric tensor into its irreducible components is detailed
in Appendix.~\ref{sec:decomp},.

\subsection{The relaxation models}

\label{sec:relaxation models}

Instead of linearization, another class of phenomenological collision model was
developed.  Instead of carrying out the collision integrals, the distribution is
simply modeled as a relaxation towards a target distribution.  Among them, the
most widely used is the celebrated single-relaxation-time, \textit{a.k.a.}\
Krook or BGK, model~\cite{Bhatnagar1954}, where the target is simply the
equilibrium
\begin{align}
	\label{eq:BGK}
	\Omega_{BGK}=-\frac 1\tau\left(f-f^{eq}\right)
\end{align}
Although retaining most of the essence of BCO in spite of its simplicity, a
notable shortcoming of this model is that all moments are relaxed towards their
equilibrium values at the same rate, resulting in a unphysical unity Prandtl
number among other deviations from reality.

Two well-known remedies exist in the literature.  First,
Holway~\cite{Holway1966} proposed the ellipsoidal statistical (ES) model which
replaces $f^{eq}$ in Eq.~\eqref{eq:BGK} with
\begin{equation}
	f_{ES} = \frac{\rho}{\det(\bm{\Lambda})^{1/2}}
	\exp\left[-\frac{\bc\bc:\bm{\Lambda}^{-1}}{2}\right] \label{eq:ES-BGK}
\end{equation}
where $\bm{\Lambda} = \theta\bdelta+b\bm{\sigma}/\rho$, and $\bm{\Lambda}^{-1}$
its tensor inverse such that $\bm{\Lambda}\cdot\bm{\Lambda}^{-1} = \bdelta$. The
Prandtl number turns out to depend on the parameter $b$ and becomes $2/3$ at
$b=-0.5$.  Another popular model is given by Shakhov~\cite{Shakhov1968} which
replaces $f^{eq}$ in Eq.~\eqref{eq:BGK} with
\begin{align}
	\label{eq:shakhov}
	f_S = f^{eq}\left[1+\left(1-\pr\right)\frac{\bm{q}\cdot\bc}{5\rho\theta^{2}}
	\left(\frac{c^2}{\theta}-5\right) \right]
\end{align}
where $\pr$ the desired
Prandtl number.  Noticing that the expression multiplied by the factor $1-\pr$
is in fact the eigen-function corresponding to heat flux, the effect of
Eq.~\eqref{eq:shakhov} is simply multiplying the eigenvalue by $\pr$ and leaving
everything else the same as in the BGK model.

\subsection{Ikenberry's spherical moment system}

Using the solid spherical harmonics, $\bY{n}(\bc)$, Ikenberry and
Truesdell~\cite{Ikenberry1956,Truesdell1956} introduced the functions
$\bY{n,m}(\bm{c}) \equiv c^{2m}\bY{n}(\bc)$ and with which the \textit{spherical
	moments} as
\begin{equation}
	\bm{P}^{(n,m)}=\int f \bY{n,m}(\bc) d\bc
\end{equation}
It can be recognized that $\bY{n,m}(\bc)$, linearly independent but necessarily
orthogonal to each other, are linear combinations of $\bdelta^m\bH{n,m}(\bc)$
and thus complete. The leading spherical moments are related to macroscopic
hydrodynamics as $\bm{P}^{(0,0)}=\rho$, $\bm{P}^{(1,0)}=0$, $\bm{P}^{(0,1)}=3p$
which are conserved, and $\bm{P}^{(2,0)}=\bm{\sigma}$, $\bm{P}^{(1,1)}=2\bm{q}$
which are non-conserved. Inspired by the thoughts of Maxwell that for Maxwell
molecules, the collision integral for a distribution in the form of a degree-$n$
polynomial can be expressed as a polynomial in the moments of degrees not
greater than $n$, Ikenberry \textit{et al} conjectured and proved the
\textit{Ikenberry's theorem} to evaluate the spherical moments of BCO for
Maxwell molecules~\cite{Ikenberry1956,Truesdell1956,Truesdell1980a}:
\begin{equation}
    \int\Omega\bY{n,m}(\bc)d\bc = -D_{n,m}\bm{P}^{(n,m)} +
    \sum D_{n_1,n_2,m_1,m_2}\bm{P}^{(n_1,m_1)}\bm{P}^{(n_2,m_2)}
\end{equation}
where $D_{0,0}$ and $D_{1,0}$ vanish due to mass and momentum conservation,
$D_{2,0}$ and $D_{1,1}$ are related to shear viscosity and thermal diffusivity
respectively.  Although formally similar to the eigen-system for linearized BCO,
the collision integral was not linearized here and hence the bilinear form on
the \textit{r.h.s.}\ of which the coefficients, $D_{n_1,n_2,m_1,m_2}$, are
non-zero only if $2m_1+n_1+ 2m_2+n_2=2m+n$. It should be emphasized that the
spherical moments are those of the distribution itself rather than its
perturbation.

\section{The spectral multiple-relaxation-time LBM model}

\label{sec:smrt}

Although the LBM was originally developed from the LGA rather independent of
classic kinetic theory, it was later formulated as a velocity-space
discretization of the Boltzmann equation with BGK collision term based on the
absolute Hermite expansion given by
Eq.~\eqref{eq:hermite1}~\cite{Shan1998,Shan2006b}.  As the early purpose of LBM
was to recover the equations for hydrodynamic variables which are low-order
velocity moments, the expansion was truncated while preserving the dynamics of
these low-order moments~\cite{Nie2008a}.  It was further noticed that for a
finite Hermite expansion, Eq.~\eqref{eq:a} can be calculated from values of the
distribution at a set of discrete velocities which form a Gauss-Hermite
quadrature rule in velocity space. Namely, letting $w_i$ and $\bxi_i$, $i = 1,
\cdots, d$, be respectively the weights and abscissas of a degree-$2N$
Gauss-Hermite quadrature, the expansion coefficients can be obtained as a
weighted sum
\begin{equation}
	\label{eq:an}
	\ba{n}(\bx,t) = \int f_N\bH{n}(\bxi)d\bxi
	= \int\omega(\bxi)\left[\frac{f_N\bH{n}(\bxi)}{\omega(\bxi)}\right]d\bxi
	= \sum_{i=1}^df_i(\bx, t)\bH{n}(\bxi_i)
\end{equation}
where $f_N = f_N(\bx, \bxi, t)$ is the order-$N$ truncation of
Eq.~\eqref{eq:hermite1}, and
\begin{equation}
	\label{eq:fi}
	f_i(\bx, t) \equiv \frac{w_if_N(\bx, \bxi_i, t)}{\omega(\bxi_i)}
	= w_i\sum_{n=0}^N\ba{n}(\bx, t):\bH{n}(\bxi_i)
\end{equation}
The last equality of Eq.~\eqref{eq:an} holds because the term in the brackets is
a polynomial of a degree not exceeding $2N$. Eqs.~\eqref{eq:an} and
\eqref{eq:fi} define an \textit{isomorphism} between $\ba{n}(\bx, t)$ and
$f_i(\bx, t)$. The dynamic equations of $f_i$ can be obtained by projecting the
Boltzmann-BGK equation into the Hilbert space spanned by the leading Hermite
polynomials and evaluating at $\bxi_i$:
\begin{equation}
	\pp{f_i}t + \bxi_i\cdot\nabla f_i = -\frac 1\tau\left(f_i - f^{eq}_i\right)
\end{equation}
where $f^{eq}_i$ is the truncated Hermite expansion of the Maxwellian evaluated
at $\bxi_i$.  After discretizing space and time, the popular LBGK schemes are
arrived at~\cite{Shan2006b}.  Two remarks can be made at this point. First, the
expansion about absolute Maxwellian instead of the local one is used because it
results in constant $\bxi_i$ which, when aligned with the underlying lattice,
yields the computationally efficient stream-collide LB algorithm.  Second,
unlike the original LBGK models obtained \textit{a posteriori} by enforcing
compliance with near-incompressible Navier-Stokes equations, the kinetic
theoretic derivation results in thermal compressible hydrodynamics and those
beyond, essentially providing a potential approach of discretizing the Boltzmann
equation itself.

One of the defects that LBGK inherited from BGK model is the fixed unity Prandtl
number due to coupled viscous and thermal transport.  The
multiple-relaxation-time (MRT) LB scheme~\cite{d'Humieres1992} attempted to
solve this problem by assigning separate relaxation times to each eigen modes of
the discrete distribution. However, as the eigen-decomposition depends on the
underlying lattice, and the commonly-used lattices are not sufficient to recover
the energy equation, the issue of coupled viscous-thermal transport was not
resolved.

Based on the Hermite-expansion derivation of LBM, a scheme was proposed to
assign a separate relaxation time to each Hermite term~\cite{Shan2007}. Although
a variable Prandtl number was successfully obtained, the thermal transport was
found to depend on the Mach number when the thermal and viscous relaxation rates
are different. This abnormality was found to be caused by the fact that the
relaxations, essentially local physics, were incorrectly associated with the
\textit{raw} moments in the absolute frame.  By converting \textit{via} binomial
transform the raw moments to the \textit{central} moments in local frame before
relaxing them, a Galilean invariant multiple-relaxation-time LB collision model
in Hermite spectral space (SMRT) was obtained~\cite{Shan2019,Li2019}.  Further
realizing that the minimum sets of components of the tensorial moments that can
be relaxed together without violating rotation symmetry correspond to the
irreducible tensors~\cite{Li2020b}, we arrived at the SMRT model~\cite{Shan2021}
which is defined \textit{via} the local Hermite expansion~\cite{Shan2021}
\begin{equation}
	f(\bx, \bv, t) = \omega(\bv)\sum_{n=0}^\infty\frac 1{n!}
	\bd{n}(\bx, t):\bH{n}(\bv)
\end{equation}
The above is otherwise identical to Eq.~\eqref{eq:hermite} except that the
factor $\rho\theta^{-3/2}$ is now absorbed into the expansion coefficients,
$\bd{n}$, which are related to the coefficients of the absolute
expansion~\eqref{eq:hermite1} \textit{via} binomial transform and \textit{vice
	versa}~\cite{Shan2021}. After decomposing $\bH{n}$ into its irreducible
components, we can write
\begin{equation}
	f(\bx, \bv, t) = \omega(\bv)\sum_{n=0}^\infty\frac{1}{n!}
	\sum_{m=0}^{\lfloor n/2\rfloor}\bd{n,m}(\bx,t):\bH{n,m}(\bv)
\end{equation}
Denoting by $\bd{n,m}_{eq}$ and $\bd{n,m}_\Omega$ the expansion coefficients of
the Maxwellian and collision term respectively, and noting that $\bd{0,0}_{eq} =
\rho$ and $\bd{n,m}_{eq} = 0$ for all $n > 0$, the SMRT operator is defined by
\begin{equation}
	\label{eq:SMRT}
	\bd{n,m}_\Omega = \left\{
	\begin{array}{ll}
		0, & n = 0, 1 \\
		-\dfrac{1}{\tau_{n,m}}\left[\bd{n,m}-\bd{n,m}_{eq}\right] =
		-\dfrac{\bd{n,m}}{\tau_{n,m}}, & n \geq 2
	\end{array}\right.
\end{equation}
where $\tau_{n,m}$ is the relaxation time of the mode $\bH{n,m}$. Particularly,
$\tau_{2,0}$ and $\tau_{2,1}$ dictate respectively the shear and bulk
viscosity of which the latter vanishes in a monatomic gas, and $\tau_{3,1}$
gives the thermal diffusivity.  The other relaxation times, although potentially
significant in non-equilibrium flows~\cite{Shi2021}, have no hydrodynamic
correspondence.

The relations between SMRT and the relaxation models can be easily identified.
First of all, when all collision frequencies are the same, Eq.~\eqref{eq:SMRT}
reduces to the BGK model regardless the frame in which the expansion is carried
out.  Noticing that the second term on the \textit{r.h.s.}\ of
Eq.~\eqref{eq:shakhov} is actually $\frac{1}{3!}(1-\pr)\bd{3,1}:\bH{3,1}$, it
can be recognized that the Shakhov model is the special case of
Eq.~\eqref{eq:SMRT} with $\tau_{3,1} = \tau_{2,0}/\pr$, and
$\tau_{n,m} = \tau_{2,0}$ for all others.  For the ES model, we point out
that $\bm{\Lambda}$ is nothing but the weighted sum of two second-order
irreducible tensors with $b$ being the proportional constant.

With $\bH{n,m}$ being the eigen-functions of the linearized BCO for Maxwell
molecules, the SMRT is somewhat theoretically justifiable by the rationale that
$\bH{n,m}$ can serve as a good basis to approximate the BCO in general as
suggested by Grad~\cite{Grad1958} and others.  In the rest of the paper we shall
numerically examine the performance of SMRT in the highly non-equilibrium
problem of shock structure.

\section{Numerical Implementation and Results}

\label{sec:numerical}

The profile of a normal shock is a typical non-linear and non-equilibrium
benchmark where the variation of macroscopic properties across the shock region
is steep, and the degree of non-equilibrium increases with the Mach number.  To
benchmark the ability of SMRT-LB in simulating non-equilibrium flows, we compute
the profile of an one-dimensional (1D) shock in a monatomic gas for Mach numbers
ranging from 1.5 to 7.

\subsection{Convergence of Hermite expansion and the shifted-stencil}

\label{sec:conv}

By the theory of generalized Fourier expansion, any square-integrable function
in the sense that $\int\omega f^2d\bxi$ exists can be expanded as the Hermite
series
\begin{equation}
	f = \sum_{n=0}^\infty\frac{1}{n!}\ba{n}:\bH{n}(\bxi)
\end{equation}
The condition for expansion~\eqref{eq:hermite1} to converge calls for the following
integral to exist:
\begin{equation}
	\int\omega\left[\frac{f(\bxi)}{\omega(\bxi)}\right]^2d\bxi =
	\int\exp\left(\frac{\xi^2}2\right)f^2(\bxi)d\bxi
\end{equation}
Assuming that $f\sim f^{eq}$ as $\xi\rightarrow\infty$, the convergence
condition is hence $\theta < 2$, \textit{i.e.}, the highest temperature in the
domain cannot exceed twice of the reference temperature. Note that in certain
moment-method calculations~\cite{Grad1952,Holway1964,Cai2019}, the reference
temperature was chosen as the upper-stream (colder) temperature which leads to a
restriction on the down-stream (hotter) temperature and the maximum achievable
Mach number in turn.  In the present work, the reference temperature was not
tied to any physical temperature but chosen to ensure that the highest
dimensionless temperature is always $<2$.

Another hidden parameter implicitly dictated by $\omega(\bxi)$ is the velocity
of the reference frame.  Galilean invariance principle implies that moments in
all inertial frames are equal in representing the macroscopic hydrodynamics --
they are in fact related to each other \textit{via} binomial
transform~\cite{Shan2021}. However, as shown in Appendix~\ref{sec:conver}, the
Hermite expansion converges faster if the distribution is closer to
$\omega(\bxi)$. In the present work, we shift the quadrature abscissas,
$\bxi_i$, by a constant $\bar{\bu} = c_l\bm{k}$ where $c_l$ is the lattice
constant of the quadrature~\cite{Shan2016} and $\bm{k}$ a constant integer
vector chosen to minimize the maximum velocity in the frame moving with
$\bar{\bu}$. Obviously, with the same weights $w_i$, the abscissas
$\bar{\bu}+\bxi_i$ constitute a quadrature rule of the same order in the moving
frame, hence the corresponding discrete distributions are isomorphic to the
moments therein and can be used as the state variable in LB. We note that the
shifted-stencil technique was also employed in Ref.~\cite{Frapolli2016} with a
more involved argument.

\subsection{Problem Statement and Implementation Details}

For a normal shock, the ratios between the up- and down-stream density, $\rho_1$
and $\rho_2$, fluid velocity relative to the shock, $u_1$ and $u_2$, and
temperature, $\theta_1$ and $\theta_2$, are given by the Rankine-Hugoniot
relations~\cite{Landau1987}
\begin{equation}
	\label{eq:rankine}
	\frac{\rho_2}{\rho_1} = \frac{u_1}{u_2} =
	\frac{(\gamma+1)\ma^2}{(\gamma-1)\ma^2+2},
	\quad\mbox{and}\quad
	\frac{\theta_2}{\theta_1} = \frac{\left(2\gamma\ma^2 - \gamma + 1\right)
		\left[(\gamma-1)\ma^2 + 2\right]}{(\gamma+1)^2\ma}
\end{equation}
where $\gamma$ is the ratio of specific heats, $\ma\equiv u_1/c_1$ and $c_1
\equiv \sqrt{\gamma\theta_1}$ are respectively the up-stream Mach number and
speed of sound. Moreover, the following relations can be obtained
\begin{equation}
	\frac{u_1 + u_2}2 = \frac{\gamma\ma^2+1}{(\gamma+1)\ma}c_1,
	\quad\mbox{and}\quad
	\frac{u_1 - u_2}2 = \frac{\ma^2-1}{(\gamma+1)\ma}c_1
\end{equation}
To accelerate the convergence of the Hermite expansion, the discrete velocities
were shifted by a $\bar{u}_x$ closest to $(u_1+u_2)/2$ to minimize the maximum
relative speed which is bounded from below at $(u_1-u_2)/2$.  The choice of
$\theta_2$ (and $\theta_1$ in turn \textit{via} Eq.~\eqref{eq:rankine}) is made
to ensure overall convergence.  As only monatomic gases are considered, $\gamma
= 5/3$.  The specific values used in all cases are listed in
Table~\ref{tab:Implementation}.  All computations were carried out on a
$100\times 1\times 1$ grid with Maxwellian specified at both ends in the
$x$-direction and periodic boundary condition used in the $y$- and
$z$-directions.  Initially the distribution was the Maxwellian with macroscopic
quantities set as step functions between the theoretical up- and down-stream
values.  To investigate the effect of the truncation order, computations were
carried out for both $N=4$ and $N=5$ using otherwise identical parameters.
Three-dimensional tensor products of the 1D quadrature rule
D1Qd-P1~\cite{Shi2021a} were used with degrees of precision, $Q$, sufficient for
the truncation.  Both $d$ and $Q$ are listed in Table~\ref{tab:Implementation}.

\begin{table}[hbt!]
	\caption{\label{tab:Implementation} Parameters used in all simulations. The
		up-stream density, $\rho_1$, is always unity.}
	\centering
	\begin{tabular}{lccccccccc}
		\hline
		$\ma$ & $d$  & $Q$ &   $c_l$   & $\rho_2$ & $\theta_1$ & $\theta_2$ & $k_x$ & $\bar{u}_x$ & $(u_1+u_2)/2$ \\ \hline
		1.5   & $9$  & 11  & 0.6780039 &   1.71   &   0.714    &    1.07    &   2   &    1.36     &     1.29      \\
		3     & $9$  & 11  & 0.6780039 &   3.00   &   0.333    &    1.22    &   2   &    1.36     &     1.49      \\
		5     & $9$  & 11  & 0.6780039 &   3.57   &   0.167    &    1.45    &   2   &    1.36     &     1.68      \\
		7     & $17$ & 19  & 0.5638701 &   3.77   &   0.071    &    1.15    &   3   &    1.69     &     1.53      \\ \hline
	\end{tabular}
\end{table}

The relaxation times are taken to be those of the Maxwell molecules as given
previously~\cite{Shi2021} and documented in Table~\ref{tab:relaxation times} as
ratios to $\tau_{2,0}$, which is specified \textit{via} the up-stream Knudsen
number $\kn = \nu / \sqrt{\theta_1} = \tau_{2,0} / \sqrt{\theta_1}$, where $\nu$
is the dimensionless kinematic viscosity. The viscosity-temperature index of
Maxwell molecules is 1. Throughout the paper, $\kn = 1$ is used when comparing
results with other methods.  We note that $\tau_{2,1}$ is absent for monatomic
gases.  As mentioned in Section~\ref{sec:relaxation models}, the Shakhov model
is equivalent to setting $\tau^\ast_{3,1} = 3/2$ and all others unity.

\newcommand{\tns}[1]{\tau^\ast_{n,#1}}

\begin{table}[hbt!]
	\caption{\label{tab:relaxation times} Relaxation times for Maxwellian molecules
	with $\tau^{*}_{n,m}=\tau_{n,m}/\tau_{2,0}$}
	\centering
	\begin{tabular}{l|ccc|ccc}
		\hline
		    &     \multicolumn{3}{c|}{SMRT}      &    \multicolumn{3}{c}{Shakhov}     \\
		$n$ & $\tns{0} $ & $\tns{1}$ & $\tns{2}$ & $\tns{0} $ & $\tns{1}$ & $\tns{2}$ \\ \hline
		2   &     1      &           &           &     1      &           &           \\
		3   &    2/3     &    3/2    &           &     1      &    3/2    &           \\
		4   &  432/763   &    6/7    &    3/2    &     1      &     1     &     1     \\
		5   &  864/1655  &  108/163  &     1     &     1      &     1     &     1     \\ \hline
	\end{tabular}
\end{table}

As the reference solution, the full Boltzmann equation was solved for Maxwell
molecules using fast spectral method (FSM)~\cite{Wu2013} on a $200\times 1\times
1$ spatial grid where, for $\ma = 7$, the velocity domain $[-24,24]^3$ was
uniformly discretized by a $168\times84\times84$ grid. For smaller Mach numbers,
smaller velocity domains and fewer discrete velocities were used.

\subsection{Results and Discussions}

\pgfplotsset{
	width=.475*\textwidth,
	height=.28*\textheight,
	cycle list={
		{black, only marks, mark=o,mark size=1pt, mark repeat=6},
		{red, thick},
		{blue, thick, dashed},
		{blue, thick, dotted}
	},
	legend cell align = {left},
	legend pos=north west,
	grid=both
}

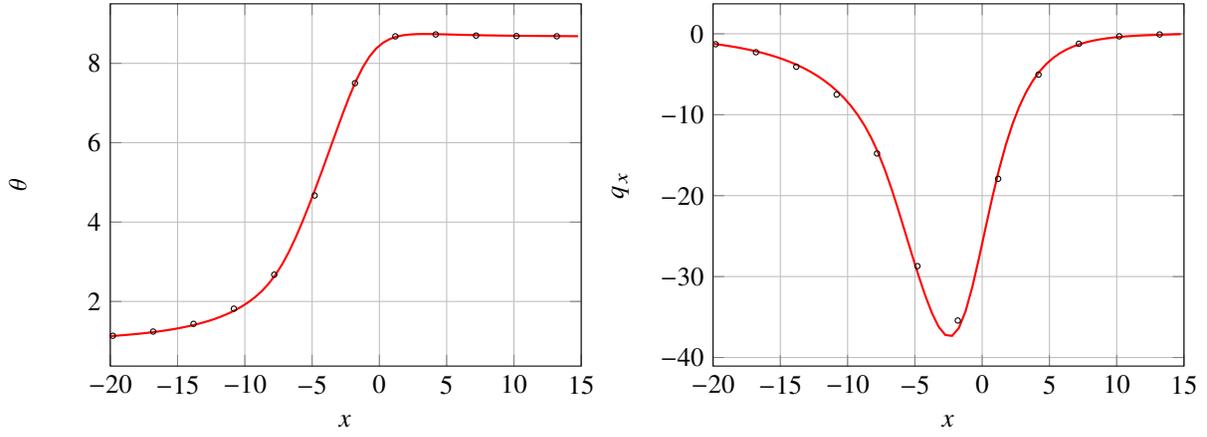
\begin{figure}[htbp]
	\centering
	\begin{tikzpicture}
		\pgfplotsset{
			xmin=-20,
			xmax=15,
			xlabel={$x$}
		}
		\matrix {
			\begin{axis}[ylabel={$\theta$}]
				\addplot table {Fig/Shock/Ma5.0/S-Model/S_Ma5.0_T.dat};
				\addplot table {Fig/Shock/DVM/Newton-Cotes/DVM_Ma5.0_NewtonCotes_T.dat};
			\end{axis}
			&
			\begin{axis}[ylabel={$q_x$}]
				\addplot table {Fig/Shock/Ma5.0/S-Model/S_Ma5.0_Q.dat}; \label{plots:LBM}
				\addplot table {Fig/Shock/DVM/Newton-Cotes/DVM_Ma5.0_NewtonCotes_Q.dat}; \label{plots:DVM}
			\end{axis}
			\\
		};
	\end{tikzpicture}
	\caption{Comparison of LBM (\ref{plots:LBM}) and DVM (\ref{plots:DVM}) at
	$\ma=5$ using Shakhov model.}
	\label{fig:lbm v.s. dvm}
\end{figure}

Firstly as a sanity check, we validate our results against a high-resolution DVM
computation of Shakhov model using Newton-Cotes
quadrature~\cite{Yang2016a} where the velocity space $[-15,15]$ was
discretized by 401 data points.  Shown in Fig.~\ref{fig:lbm v.s. dvm} are
temperature and heat flux at $\ma = 5$.  Density and velocity results are
omitted as there is no distinguishable difference between the two methods.  It
can be seen that the results of LBM and DVM are in good agreement with a
relative error of no more than 3\%.  It is worth noting that the LBM computation
used only nine discrete velocities in each direction while capturing the highly
non-equilibrium effect reasonably well at a high Mach number.

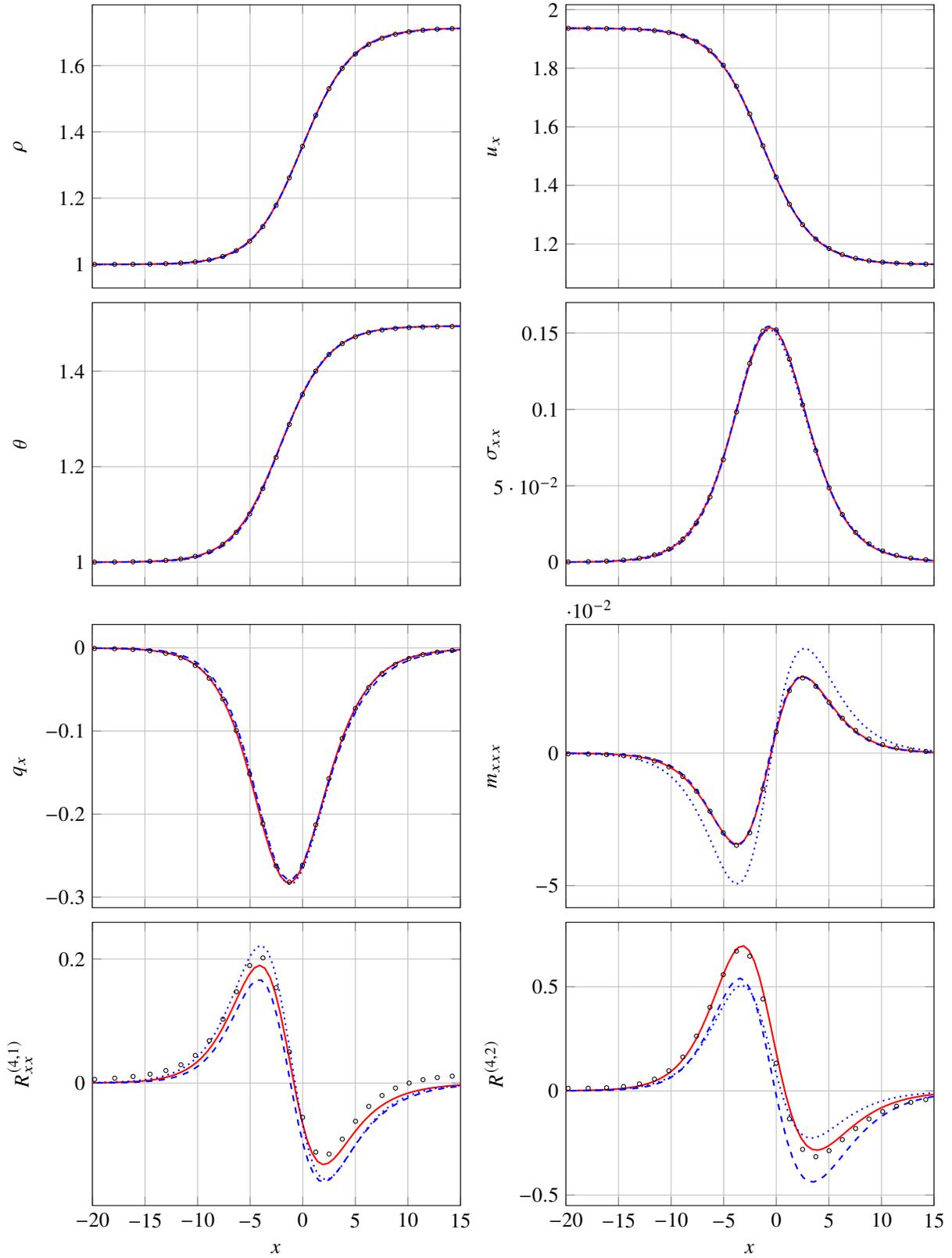
\begin{figure}
	\centering
	\begin{tikzpicture}
		\pgfplotsset{
			xmin=-20,
			xmax=15
		}
		\matrix {
			\begin{axis}[xticklabel=\empty,ylabel={$\rho$}]
				\addplot table{Fig/Shock/Ma1.5/FSM/FSM_Ma1.5_Rho.dat}; \label{plots:Boltzmann}
				\addplot table{Fig/Shock/Ma1.5/SMRT/SMRT_Ma1.5_Rho.dat}; \label{plots:SMRT5}
				\addplot table{Fig/Shock/N4th/Ma1.5/4th_1.5_Rho.dat}; \label{plots:SMRT4}
				\addplot table{Fig/Shock/Ma1.5/S-Model/S_Ma1.5_Rho.dat}; \label{plots:s-model}
			\end{axis}
			&
			\begin{axis}[xticklabel=\empty,,ylabel={$u_x$}]
				\addplot table{Fig/Shock/Ma1.5/FSM/FSM_Ma1.5_U.dat};
				\addplot table{Fig/Shock/Ma1.5/SMRT/SMRT_Ma1.5_U.dat};
				\addplot table{Fig/Shock/N4th/Ma1.5/4th_1.5_U.dat};
				\addplot table{Fig/Shock/Ma1.5/S-Model/S_Ma1.5_U.dat};
			\end{axis}
			\\
			\begin{axis}[xticklabel=\empty,ylabel={$\theta$}]
				\addplot table{Fig/Shock/Ma1.5/FSM/FSM_Ma1.5_T.dat};
				\addplot table{Fig/Shock/Ma1.5/SMRT/SMRT_Ma1.5_T.dat};
				\addplot table{Fig/Shock/N4th/Ma1.5/4th_1.5_T.dat};
				\addplot table{Fig/Shock/Ma1.5/S-Model/S_Ma1.5_T.dat};
			\end{axis}
			&
			\begin{axis}[xticklabel=\empty,,ylabel={$\sigma_{xx}$}]
				\addplot table{Fig/Shock/Ma1.5/FSM/FSM_Ma1.5_S.dat};
				\addplot table{Fig/Shock/Ma1.5/SMRT/SMRT_Ma1.5_S.dat};
				\addplot table{Fig/Shock/N4th/Ma1.5/4th_1.5_S.dat};
				\addplot table{Fig/Shock/Ma1.5/S-Model/S_Ma1.5_S.dat};
			\end{axis}
			\\
			\begin{axis}[xticklabel=\empty,ylabel={$q_x$}]
				\addplot table{Fig/Shock/Ma1.5/FSM/FSM_Ma1.5_Q.dat};
				\addplot table{Fig/Shock/Ma1.5/SMRT/SMRT_Ma1.5_Q.dat};
				\addplot table{Fig/Shock/N4th/Ma1.5/4th_1.5_Q.dat};
				\addplot table{Fig/Shock/Ma1.5/S-Model/S_Ma1.5_Q.dat};
			\end{axis}
			&
			\begin{axis}[xticklabel=\empty,,ylabel={$m_{xxx}$}]
				\addplot table{Fig/Shock/Ma1.5/FSM/FSM_Ma1.5_M.dat};
				\addplot table{Fig/Shock/Ma1.5/SMRT/SMRT_Ma1.5_M.dat};
				\addplot table{Fig/Shock/N4th/Ma1.5/4th_1.5_M.dat};
				\addplot table{Fig/Shock/Ma1.5/S-Model/S_Ma1.5_M.dat};
			\end{axis}
			\\
			\begin{axis}[xlabel={$x$},ylabel={$R^{(4,1)}_{xx}$}]
				\addplot table{Fig/Shock/Ma1.5/FSM/FSM_Ma1.5_R2.dat};
				\addplot table{Fig/Shock/Ma1.5/SMRT/SMRT_Ma1.5_R2.dat};
				\addplot table{Fig/Shock/N4th/Ma1.5/4th_1.5_R2.dat};
				\addplot table{Fig/Shock/Ma1.5/S-Model/S_Ma1.5_R2.dat};
			\end{axis}
			&
			\begin{axis}[xlabel={$x$},ylabel={$R^{(4,2)}$}]
				\addplot table{Fig/Shock/Ma1.5/FSM/FSM_Ma1.5_R0.dat};
				\addplot table{Fig/Shock/Ma1.5/SMRT/SMRT_Ma1.5_R0.dat};
				\addplot table{Fig/Shock/N4th/Ma1.5/4th_1.5_R0.dat};
				\addplot table{Fig/Shock/Ma1.5/S-Model/S_Ma1.5_R0.dat};
			\end{axis}
			\\
		};
	\end{tikzpicture}
	\caption{Shock profiles at $\ma=1.5$ as computed by the 4-th
	(\ref{plots:SMRT4}) and 5-th (\ref{plots:SMRT5}) order SMRT model, the Shakhov
	model (\ref{plots:s-model}) as defined in Table~\ref{tab:relaxation times}, and
	by the reference  fast spectral method (\ref{plots:Boltzmann}).}
	\label{fig:profile}
\end{figure}

\begin{figure}
	\centering
	\begin{tikzpicture}
		\pgfplotsset{
			xmin=-20,
			xmax=15
		}
		\matrix {
			\begin{axis}[xticklabel=\empty,ylabel={$\rho$}]
				\addplot table{Fig/Shock/Ma3.0/FSM/FSM_Ma3.0_Rho.dat};
				\addplot table{Fig/Shock/Ma3.0/SMRT/SMRT_Ma3.0_Rho.dat};
				\addplot table{Fig/Shock/N4th/Ma3.0/4th_3.0_Rho.dat};
				\addplot table{Fig/Shock/Ma3.0/S-Model/S_Ma3.0_Rho.dat};
			\end{axis}
			&
			\begin{axis}[xticklabel=\empty,,ylabel={$u_x$}]
				\addplot table{Fig/Shock/Ma3.0/FSM/FSM_Ma3.0_U.dat};
				\addplot table{Fig/Shock/Ma3.0/SMRT/SMRT_Ma3.0_U.dat};
				\addplot table{Fig/Shock/N4th/Ma3.0/4th_3.0_U.dat};
				\addplot table{Fig/Shock/Ma3.0/S-Model/S_Ma3.0_U.dat};
			\end{axis}
			\\
			\begin{axis}[xticklabel=\empty,ylabel={$\theta$}]
				\addplot table{Fig/Shock/Ma3.0/FSM/FSM_Ma3.0_T.dat};
				\addplot table{Fig/Shock/Ma3.0/SMRT/SMRT_Ma3.0_T.dat};
				\addplot table{Fig/Shock/N4th/Ma3.0/4th_3.0_T.dat};
				\addplot table{Fig/Shock/Ma3.0/S-Model/S_Ma3.0_T.dat};
			\end{axis}
			&
			\begin{axis}[xticklabel=\empty,,ylabel={$\sigma_{xx}$}]
				\addplot table{Fig/Shock/Ma3.0/FSM/FSM_Ma3.0_S.dat};
				\addplot table{Fig/Shock/Ma3.0/SMRT/SMRT_Ma3.0_S.dat};
				\addplot table{Fig/Shock/N4th/Ma3.0/4th_3.0_S.dat};
				\addplot table{Fig/Shock/Ma3.0/S-Model/S_Ma3.0_S.dat};
			\end{axis}
			\\
			\begin{axis}[xticklabel=\empty,ylabel={$q_x$}]
				\addplot table{Fig/Shock/Ma3.0/FSM/FSM_Ma3.0_Q.dat};
				\addplot table{Fig/Shock/Ma3.0/SMRT/SMRT_Ma3.0_Q.dat};
				\addplot table{Fig/Shock/N4th/Ma3.0/4th_3.0_Q.dat};
				\addplot table{Fig/Shock/Ma3.0/S-Model/S_Ma3.0_Q.dat};
			\end{axis}
			&
			\begin{axis}[xticklabel=\empty,,ylabel={$m_{xxx}$}]
				\addplot table{Fig/Shock/Ma3.0/FSM/FSM_Ma3.0_M.dat};
				\addplot table{Fig/Shock/Ma3.0/SMRT/SMRT_Ma3.0_M.dat};
				\addplot table{Fig/Shock/N4th/Ma3.0/4th_3.0_M.dat};
				\addplot table{Fig/Shock/Ma3.0/S-Model/S_Ma3.0_M.dat};
			\end{axis}
			\\
			\begin{axis}[xlabel={$x$},ylabel={$R^{(4,1)}_{xx}$}]
				\addplot table{Fig/Shock/Ma3.0/FSM/FSM_Ma3.0_R2.dat};
				\addplot table{Fig/Shock/Ma3.0/SMRT/SMRT_Ma3.0_R2.dat};
				\addplot table{Fig/Shock/N4th/Ma3.0/4th_3.0_R2.dat};
				\addplot table{Fig/Shock/Ma3.0/S-Model/S_Ma3.0_R2.dat};
			\end{axis}
			&
			\begin{axis}[xlabel={$x$},ylabel={$R^{(4,2)}$}]
				\addplot table{Fig/Shock/Ma3.0/FSM/FSM_Ma3.0_R0.dat};
				\addplot table{Fig/Shock/Ma3.0/SMRT/SMRT_Ma3.0_R0.dat};
				\addplot table{Fig/Shock/N4th/Ma3.0/4th_3.0_R0.dat};
				\addplot table{Fig/Shock/Ma3.0/S-Model/S_Ma3.0_R0.dat};
			\end{axis}
			\\
		};
	\end{tikzpicture}
	\caption{Shock profiles at $\ma=3$ as computed by the 4-th
	(\ref{plots:SMRT4}) and 5-th (\ref{plots:SMRT5}) order SMRT model, the Shakhov
	model (\ref{plots:s-model}) as defined in Table~\ref{tab:relaxation times}, and
	by the reference  fast spectral method (\ref{plots:Boltzmann}).}
	\label{fig:profile3}
\end{figure}
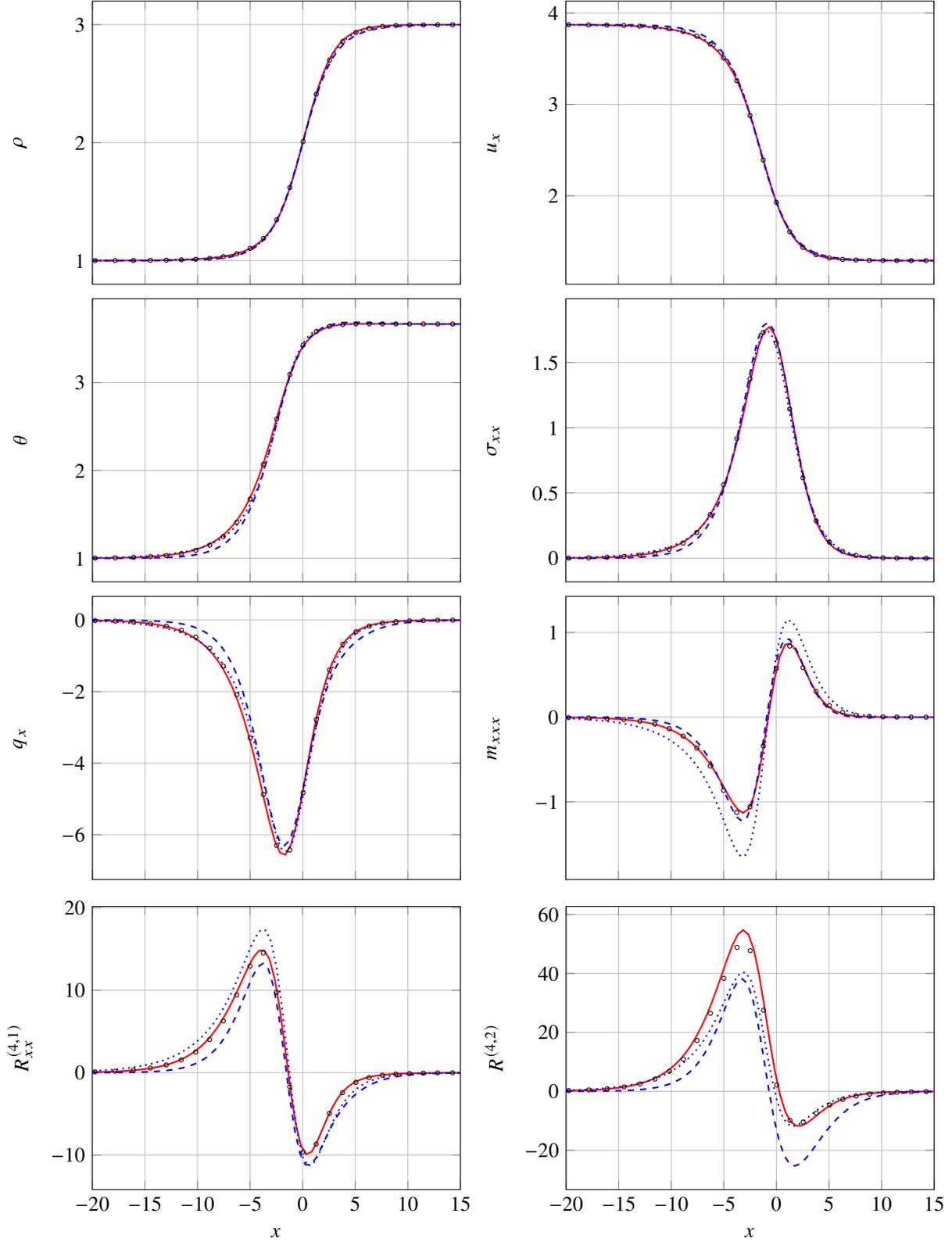

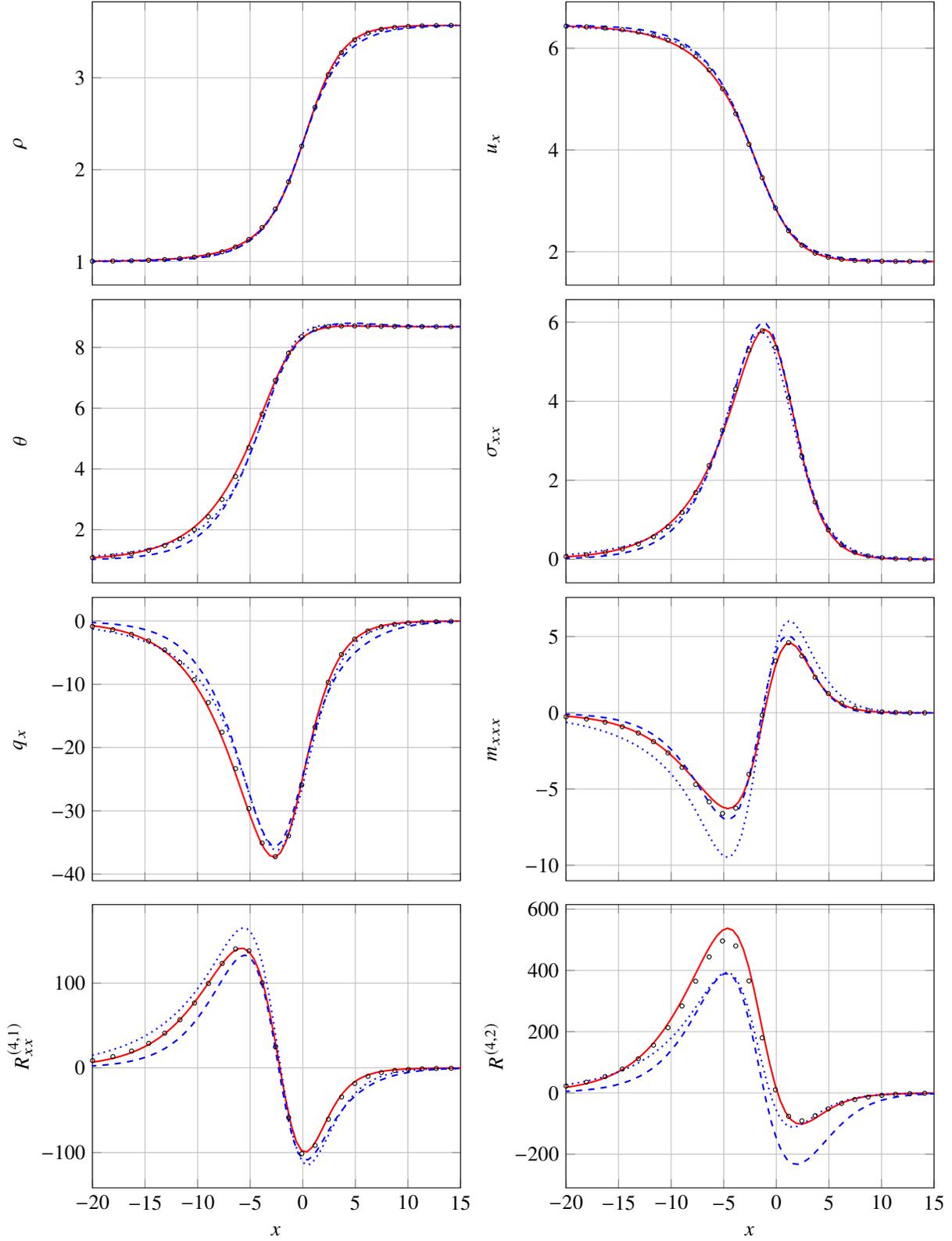
\begin{figure}
	\centering
	\begin{tikzpicture}
		\pgfplotsset{
			xmin=-20,
			xmax=15
		}
		\matrix {
			\begin{axis}[xticklabel=\empty,ylabel={$\rho$}]				
				\addplot table{Fig/Shock/Ma5.0/FSM/FSM_Ma5.0_Rho.dat};
				\addplot table{Fig/Shock/Ma5.0/SMRT/SMRT_Ma5.0_Rho.dat};
				\addplot table{Fig/Shock/N4th/Ma5.0/4th_5.0_Rho.dat};
				\addplot table{Fig/Shock/Ma5.0/S-Model/S_Ma5.0_Rho.dat};
			\end{axis}
			&
			\begin{axis}[xticklabel=\empty,,ylabel={$u_x$}]
				\addplot table{Fig/Shock/Ma5.0/FSM/FSM_Ma5.0_U.dat};
				\addplot table{Fig/Shock/Ma5.0/SMRT/SMRT_Ma5.0_U.dat};
				\addplot table{Fig/Shock/N4th/Ma5.0/4th_5.0_U.dat};
				\addplot table{Fig/Shock/Ma5.0/S-Model/S_Ma5.0_U.dat};
			\end{axis}
			\\
			\begin{axis}[xticklabel=\empty,ylabel={$\theta$}]
				\addplot table{Fig/Shock/Ma5.0/FSM/FSM_Ma5.0_T.dat};
				\addplot table{Fig/Shock/Ma5.0/SMRT/SMRT_Ma5.0_T.dat};
				\addplot table{Fig/Shock/N4th/Ma5.0/4th_5.0_T.dat};
				\addplot table{Fig/Shock/Ma5.0/S-Model/S_Ma5.0_T.dat};
			\end{axis}
			&
			\begin{axis}[xticklabel=\empty,,ylabel={$\sigma_{xx}$}]
				\addplot table{Fig/Shock/Ma5.0/FSM/FSM_Ma5.0_S.dat};
				\addplot table{Fig/Shock/Ma5.0/SMRT/SMRT_Ma5.0_S.dat};
				\addplot table{Fig/Shock/N4th/Ma5.0/4th_5.0_S.dat};
				\addplot table{Fig/Shock/Ma5.0/S-Model/S_Ma5.0_S.dat};
			\end{axis}
			\\
			\begin{axis}[xticklabel=\empty,ylabel={$q_x$}]
				\addplot table{Fig/Shock/Ma5.0/FSM/FSM_Ma5.0_Q.dat};
				\addplot table{Fig/Shock/Ma5.0/SMRT/SMRT_Ma5.0_Q.dat};
				\addplot table{Fig/Shock/N4th/Ma5.0/4th_5.0_Q.dat};
				\addplot table{Fig/Shock/Ma5.0/S-Model/S_Ma5.0_Q.dat};
			\end{axis}
			&
			\begin{axis}[xticklabel=\empty,,ylabel={$m_{xxx}$}]
				\addplot table{Fig/Shock/Ma5.0/FSM/FSM_Ma5.0_M.dat};
				\addplot table{Fig/Shock/Ma5.0/SMRT/SMRT_Ma5.0_M.dat};
				\addplot table{Fig/Shock/N4th/Ma5.0/4th_5.0_M.dat};
				\addplot table{Fig/Shock/Ma5.0/S-Model/S_Ma5.0_M.dat};
			\end{axis}
			\\
			\begin{axis}[xlabel={$x$},ylabel={$R^{(4,1)}_{xx}$}]
				\addplot table{Fig/Shock/Ma5.0/FSM/FSM_Ma5.0_R2.dat};
				\addplot table{Fig/Shock/Ma5.0/SMRT/SMRT_Ma5.0_R2.dat};
				\addplot table{Fig/Shock/N4th/Ma5.0/4th_5.0_R2.dat};
				\addplot table{Fig/Shock/Ma5.0/S-Model/S_Ma5.0_R2.dat};
			\end{axis}
			&
			\begin{axis}[xlabel={$x$},ylabel={$R^{(4,2)}$}]
				\addplot table{Fig/Shock/Ma5.0/FSM/FSM_Ma5.0_R0.dat};
				\addplot table{Fig/Shock/Ma5.0/SMRT/SMRT_Ma5.0_R0.dat};
				\addplot table{Fig/Shock/N4th/Ma5.0/4th_5.0_R0.dat};
				\addplot table{Fig/Shock/Ma5.0/S-Model/S_Ma5.0_R0.dat};
			\end{axis}
			\\
		};
	\end{tikzpicture}
	\caption{Shock profiles at $\ma=5$ as computed by the 4-th
	(\ref{plots:SMRT4}) and 5-th (\ref{plots:SMRT5}) order SMRT model, the Shakhov
	model (\ref{plots:s-model}) as defined in Table~\ref{tab:relaxation times}, and
	by the reference  fast spectral method (\ref{plots:Boltzmann}).}
	\label{fig:profile5}
\end{figure}

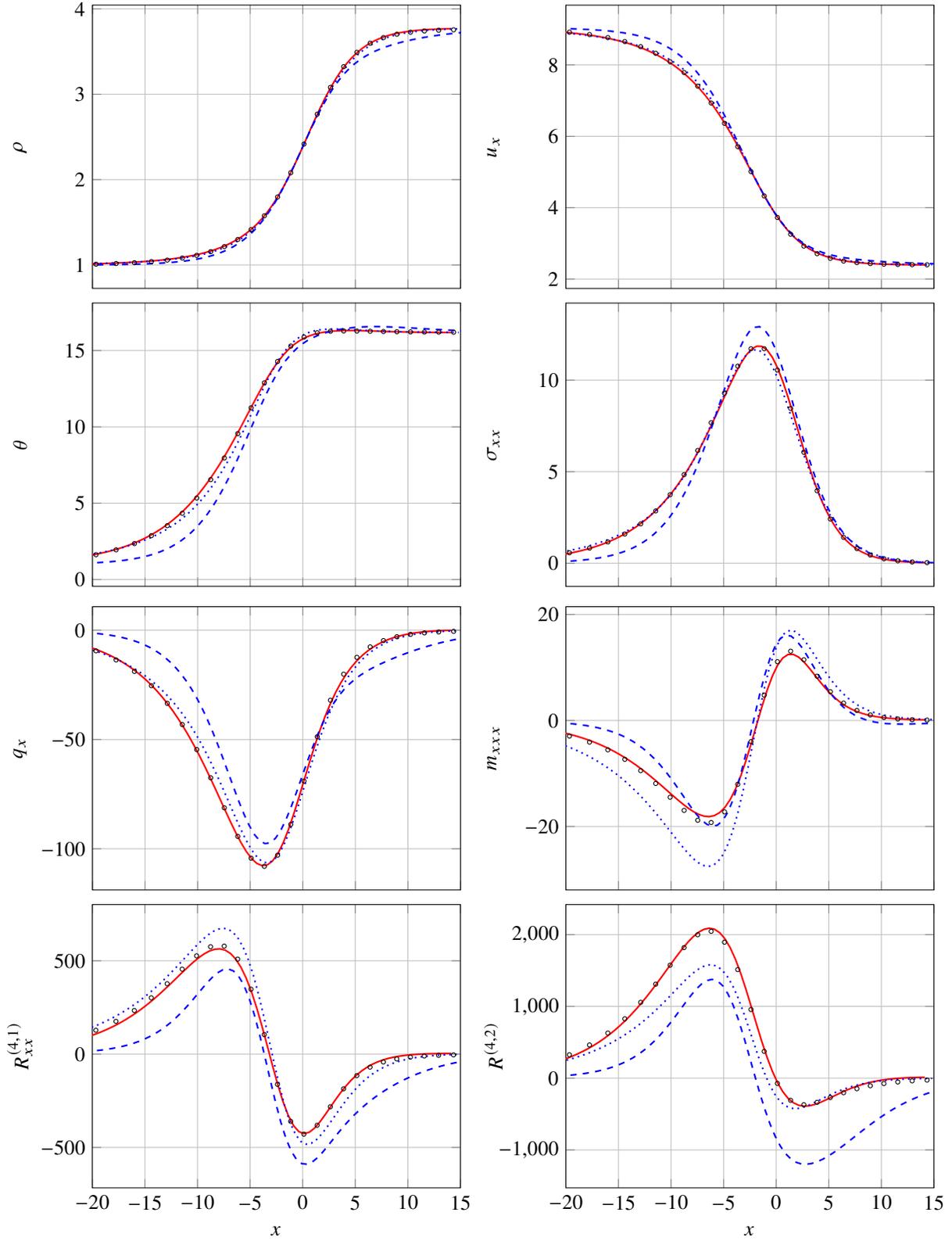
\begin{figure}
	\centering
	\begin{tikzpicture}
		\pgfplotsset{
			xmin=-20,
			xmax=15
		}
		\matrix {
			\begin{axis}[xticklabel=\empty,ylabel={$\rho$}]
				\addplot table{Fig/Shock/Ma7.0/FSM/FSM_Ma7.0_Rho.dat};
				\addplot table{Fig/Shock/Ma7.0/SMRT/SMRT_Ma7.0_Rho.dat};
				\addplot table{Fig/Shock/N4th/Ma7.0/4th_7.0_Rho.dat};
				\addplot table{Fig/Shock/Ma7.0/S-Model/S_Ma7.0_Rho.dat};
			\end{axis}
			&
			\begin{axis}[xticklabel=\empty,,ylabel={$u_x$}]
				\addplot table{Fig/Shock/Ma7.0/FSM/FSM_Ma7.0_U.dat};
				\addplot table{Fig/Shock/Ma7.0/SMRT/SMRT_Ma7.0_U.dat};
				\addplot table{Fig/Shock/N4th/Ma7.0/4th_7.0_U.dat};
				\addplot table{Fig/Shock/Ma7.0/S-Model/S_Ma7.0_U.dat};
			\end{axis}
			\\
			\begin{axis}[xticklabel=\empty,ylabel={$\theta$}]
				\addplot table{Fig/Shock/Ma7.0/FSM/FSM_Ma7.0_T.dat};
				\addplot table{Fig/Shock/Ma7.0/SMRT/SMRT_Ma7.0_T.dat};
				\addplot table{Fig/Shock/N4th/Ma7.0/4th_7.0_T.dat};
				\addplot table{Fig/Shock/Ma7.0/S-Model/S_Ma7.0_T.dat};
			\end{axis}
			&
			\begin{axis}[xticklabel=\empty,,ylabel={$\sigma_{xx}$}]
				\addplot table{Fig/Shock/Ma7.0/FSM/FSM_Ma7.0_S.dat};
				\addplot table{Fig/Shock/Ma7.0/SMRT/SMRT_Ma7.0_S.dat};
				\addplot table{Fig/Shock/N4th/Ma7.0/4th_7.0_S.dat};
				\addplot table{Fig/Shock/Ma7.0/S-Model/S_Ma7.0_S.dat};
			\end{axis}
			\\
			\begin{axis}[xticklabel=\empty,ylabel={$q_x$}]
				\addplot table{Fig/Shock/Ma7.0/FSM/FSM_Ma7.0_Q.dat};
				\addplot table{Fig/Shock/Ma7.0/SMRT/SMRT_Ma7.0_Q.dat};
				\addplot table{Fig/Shock/N4th/Ma7.0/4th_7.0_Q.dat};
				\addplot table{Fig/Shock/Ma7.0/S-Model/S_Ma7.0_Q.dat};
			\end{axis}
			&
			\begin{axis}[xticklabel=\empty,,ylabel={$m_{xxx}$}]
				\addplot table{Fig/Shock/Ma7.0/FSM/FSM_Ma7.0_M.dat};
				\addplot table{Fig/Shock/Ma7.0/SMRT/SMRT_Ma7.0_M.dat};
				\addplot table{Fig/Shock/N4th/Ma7.0/4th_7.0_M.dat};
				\addplot table{Fig/Shock/Ma7.0/S-Model/S_Ma7.0_M.dat};
			\end{axis}
			\\
			\begin{axis}[xlabel={$x$},ylabel={$R^{(4,1)}_{xx}$}]
				\addplot table{Fig/Shock/Ma7.0/FSM/FSM_Ma7.0_R2.dat};
				\addplot table{Fig/Shock/Ma7.0/SMRT/SMRT_Ma7.0_R2.dat};
				\addplot table{Fig/Shock/N4th/Ma7.0/4th_7.0_R2.dat};
				\addplot table{Fig/Shock/Ma7.0/S-Model/S_Ma7.0_R2.dat};
			\end{axis}
			&
			\begin{axis}[xlabel={$x$},ylabel={$R^{(4,2)}$}]
				\addplot table{Fig/Shock/Ma7.0/FSM/FSM_Ma7.0_R0.dat};
				\addplot table{Fig/Shock/Ma7.0/SMRT/SMRT_Ma7.0_R0.dat};
				\addplot table{Fig/Shock/N4th/Ma7.0/4th_7.0_R0.dat};
				\addplot table{Fig/Shock/Ma7.0/S-Model/S_Ma7.0_R0.dat};
			\end{axis}
			\\
		};
	\end{tikzpicture}
	\caption{Shock profiles at $\ma=7$ as computed by the 4-th
	(\ref{plots:SMRT4}) and 5-th (\ref{plots:SMRT5}) order SMRT model, the Shakhov
	model (\ref{plots:s-model}) as defined in Table~\ref{tab:relaxation times}, and
	by the reference  fast spectral method (\ref{plots:Boltzmann}).}
	\label{fig:profile7}
\end{figure}

\begin{figure}[htbp]
	\centering
	\begin{tikzpicture}
		\pgfplotsset{
			xmin=-15,
			xmax=0
		}
		\matrix {
			\begin{axis}[xticklabel=\empty,ylabel={$\theta$}]
				\addplot table{Fig/Shock/Ma7.0/FSM/FSM_Ma7.0_T.dat};
				\addplot table{Fig/Shock/Ma7.0/SMRT/SMRT_Ma7.0_T.dat};
				\addplot table{Fig/Shock/Ma7.0/S-Model/S_Ma7.0_T.dat};
			\end{axis}
			&
			\begin{axis}[xticklabel=\empty,ylabel={$q_x$}]
				\addplot table{Fig/Shock/Ma7.0/FSM/FSM_Ma7.0_Q.dat};\label{plots:q_FSM_1}
				\addplot table{Fig/Shock/Ma7.0/SMRT/SMRT_Ma7.0_Q.dat};\label{plots:q_SMRT_1}
				\addplot table{Fig/Shock/Ma7.0/S-Model/S_Ma7.0_Q.dat};\label{plots:q_S_1}
			\end{axis}
			\\
			\begin{axis}[xlabel={$x$},ylabel={$err_{\theta}(\%)$}]
				\addplot[red, only marks, mark=square, mark size=1pt]
					table{Fig/Shock/error/Kn1.0/T/error_SMRT_Ma7.0_T.dat}; \label{plots:e_SMRT_1}
				\addplot[blue, only marks, mark=o,mark size=1pt]
					table{Fig/Shock/error/Kn1.0/T/error_S_Ma7.0_T.dat}; \label{plots:e_S_1}
			\end{axis}
			&
			\begin{axis}[xlabel={$x$},ylabel={$err_{q_x}(\%)$}]
				\addplot[red,  only marks, mark=square,mark size=1pt]
					table{Fig/Shock/error/Kn1.0/Q/error_SMRT_Ma7.0_Q.dat};
				\addplot[blue, only marks, mark=o,mark size=1pt]
					table{Fig/Shock/error/Kn1.0/Q/error_S_Ma7.0_Q.dat};
			\end{axis}
			\\
		};
	\end{tikzpicture}
	\caption{Deviations of temperature and heat flux at $\ma=7$ as computed by SMRT
	and Shakhov model.  Shown on top are the results of the 5-th order SMRT
	(\ref{plots:q_SMRT_1}), the Shakhov model (\ref{plots:q_S_1}), and the
	reference FSM (\ref{plots:q_FSM_1}), and on the bottom are the relative
	percentage errors of the 5-th order SMRT (\ref{plots:e_SMRT_1}) and the Shakhov
	model (\ref{plots:e_S_1}).}

	\label{fig:error}
\end{figure}
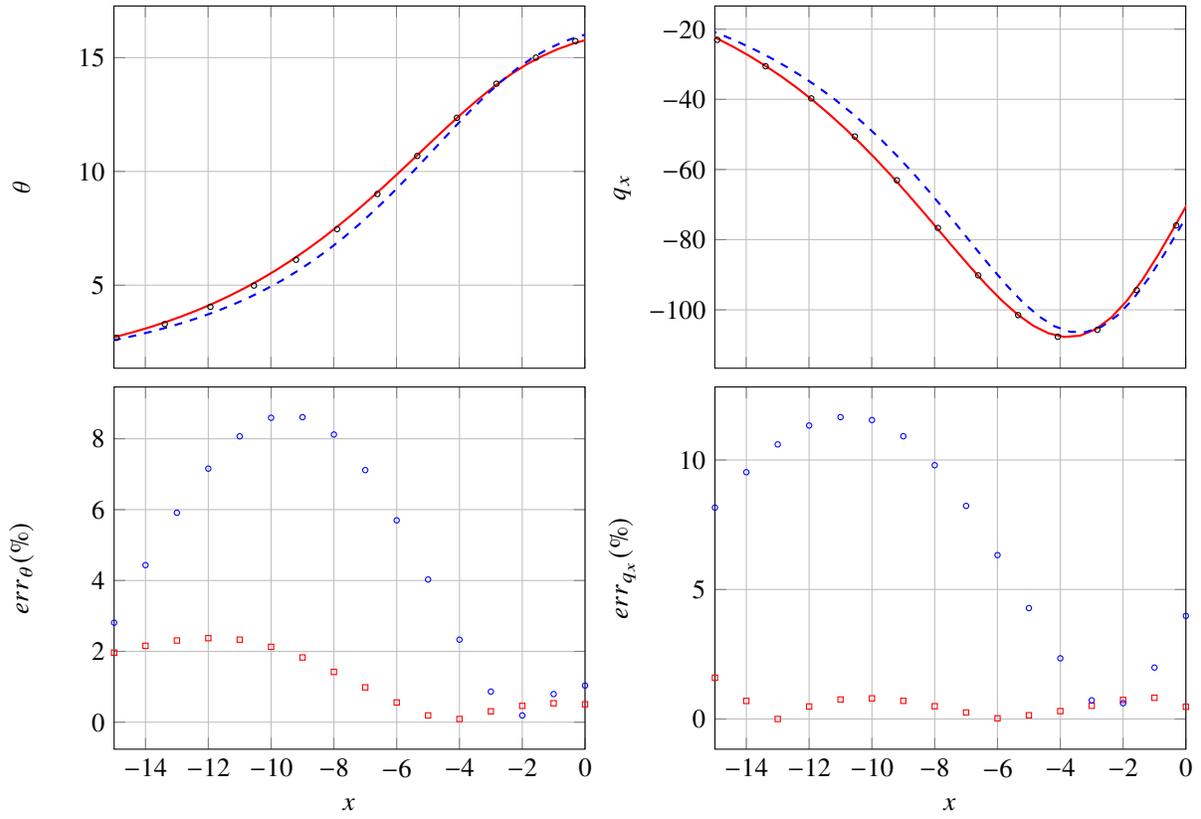

Shown in Figs.~\ref{fig:profile}--\ref{fig:profile7} are shock profiles at
$\ma=1.5$, 3, 5 and 7 as computed by the 4th- and 5th-order LBM-SMRT scheme and
by FSM.  As a comparison, results of the Shakhov model are also shown.  In
addition to density, velocity, temperature and heat flux, shock profiles of the
higher order moments: $\bm{m} = \bd{3,0} / D(3,0)$, $\bm{R}^{(4,1)} = \bd{4,1} /
D(4,1)$ and $\bm{R}^{(4,2)} = \bd{4,2} / D(4,2)$ are also shown.  Here $D(n,m)$
is given by Eq.~\eqref{eq:bd}.  A number of observations can be immediately
made.
\begin{enumerate}
	\item The fifth-order LBM gives results in excellent agreement with FSM in all
	moments up to the fourth-order for all Mach numbers tested.  We note that for
	$\ma=7$ a higher order quadrature rule was used.  Given that the number of
	velocities used here is significantly less than those normally employed in DVM
	computations, the LBM provides a very attractive alternative for computation of
	highly non-equilibrium flows.
	
	\item The fourth-order LBM yields good results only in moments up to the
	third-order and only for $\ma = 1.5$.  Deviations are visible in the fourth
	moments for all Mach numbers and starts to show up in lower moments as $\ma$
	increases. At $\ma = 7$, even the density profile becomes inaccurate.  This
	behavior seems to indicate that the truncation level of $N=5$ is necessary for
	non-equilibrium flows at high Mach numbers.
	
	\item The Shakhov model gave good results in profiles of density, velocity,
	temperature and heat flux but produced significant deviations in $m_{xxx}$ and
	all fourth-moments at all Mach numbers.  At $\ma=7$ the deviations in
	temperature and heat flux profiles become significant as displayed in
	Fig.~\ref{fig:error}.
\end{enumerate}
As an explanation to the observations above, we remark that since the transport
equation of low-order moments involves moments of higher orders, any error in
higher moments can percolate into the dynamics of the lower ones.  As the degree
of non-equilibrium increases, the higher moments become more significant, and
the hydrodynamics dictated by the dynamics of the low-order moments is more and
more affected by the error in high order moments, and accurate modeling of the
high order moments becomes more critical.

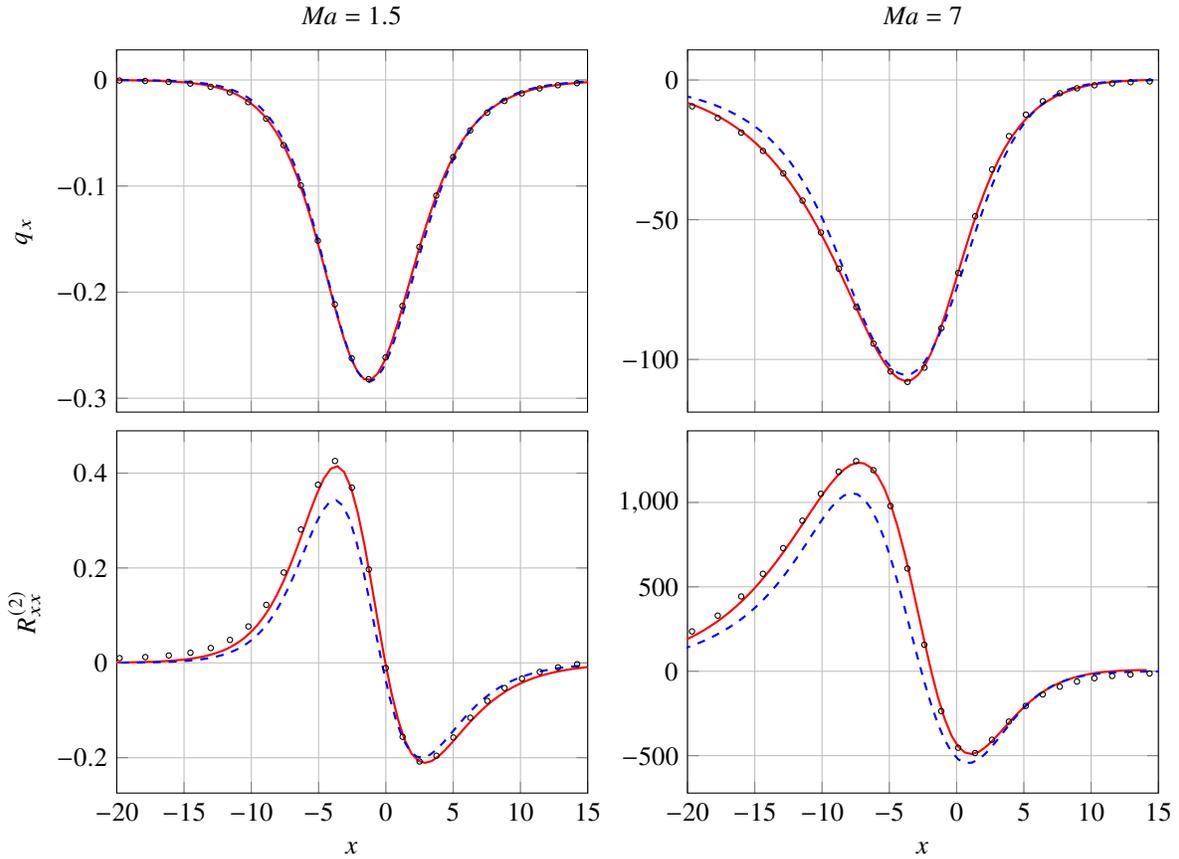
\begin{figure}[htbp]
	\centering
	\begin{tikzpicture}
		\pgfplotsset{
			xmin=-20,
			xmax=15
		}
		\matrix {
			\begin{axis}[xticklabel=\empty,ylabel={$q_x$},title={$\ma=1.5$},legend pos=south west]
				\addplot table{Fig/Shock/Ma1.5/FSM/FSM_Ma1.5_Q.dat};
				\addplot table{Fig/Shock/Ma1.5/SMRT/SMRT_Ma1.5_Q.dat};
				\addplot table{Fig/Shock/Ma1.5/tau43/Ma1.5_tau43_Q.dat};
			\end{axis}
			&
			\begin{axis}[xticklabel=\empty,title={$\ma=7$}]
				\addplot table{Fig/Shock/Ma7.0/FSM/FSM_Ma7.0_Q.dat};
				\addplot table{Fig/Shock/Ma7.0/SMRT/SMRT_Ma7.0_Q.dat};
				\addplot table{Fig/Shock/Ma7.0/tau43/Ma7.0_tau43_Q.dat};
			\end{axis}
			\\
			\begin{axis}[ylabel={$R^{(2)}_{xx}$},xlabel={$x$}]
				\addplot table{Fig/Shock/Ma1.5/FSM/FSM_Ma1.5_R.dat};
				\addplot table{Fig/Shock/Ma1.5/SMRT/SMRT_Ma1.5_R.dat};
				\addplot table{Fig/Shock/Ma1.5/tau43/Ma1.5_tau43_R.dat};
			\end{axis}
			&
			\begin{axis}[xlabel={$x$}]
				\addplot table{Fig/Shock/Ma7.0/FSM/FSM_Ma7.0_R.dat};
				\addplot table{Fig/Shock/Ma7.0/SMRT/SMRT_Ma7.0_R.dat};
				\addplot table{Fig/Shock/Ma7.0/tau43/Ma7.0_tau43_R.dat};
			\end{axis}
			\\
		};
	\end{tikzpicture}
	\caption{Comparison of SMRT (\ref{plots:SMRT5}) and reduced relaxation
	(\ref{plots:SMRT4}) with FSM (\ref{plots:Boltzmann}) at $\ma=1.5$ and $7$.}
	\label{fig:Ir or R}
\end{figure}

To illustrate the significance of allowing the irreducible components of the
Hermite polynomials to be relaxed independently, we force $\tau_{4,1} =
\tau_{4,2} = 6/7$ to demonstrate the effects of the full degree of freedom in
the relaxation.  Shown in Fig.~\ref{fig:Ir or R} are the profiles of heat flux
and the tensor $\bm{R}^{(2)} = \bm{R}^{(4,1)} + \bm{R}^{(4,2)}\bdelta$ for
$\ma=1.5$ and 7 as computed by FSM, LBM-SMRT and the ``reduced'' collision term.
To be seen is that while the error in $R^{(2)}_{xx}$ is eminent for both low and
high Mach numbers, the error in heat flux is minimum for $\ma=1.5$ but
significant for $\ma=7$.  It can be concluded that although the hydrodynamics
might not be significantly affected by the reduced relaxation in weakly
non-equilibrium flows, it is critical to relax each irreducible component
accurately and independently when the degree of non-equilibrium is strong.

\section{Conclusion}

\label{sec:conclusion}

We demonstrate that the physical foundation of the SMRT model is equivalent to
Grad's eigen-system where the eigen-state, obtained by irreducible tensor
decomposition of Hermite polynomials, is those of the linearized Boltzmann
equation for Maxwell molecules.  The LB-SMRT scheme can be viewed as a fast
discretization of moment method or a DVM scheme with the discrete velocities
chosen to ensure exact recovery of the moments.  Numerical results indicate that
retaining moments only up to the fifth order, the strongly non-equilibrium flow
inside a shock at high Mach numbers ($\ma > 7$) can be computed to satisfactory
accuracy provided that 1) the reference temperature and velocity are properly
chosen to ensure fast convergence of the Hermite expansion, and 2) all
irreducible tensor components are relaxed with the correct relaxation times. If
a simplified set of relaxation times is used, such as in the Shakhov model or
the lumped-sum Hermite polynomials relaxation, significant discrepancies can
occur, first in high-order moments and then percolating to the lower moments as
Mach number is increased.

We note that the analysis and calculation here are based on Maxwell molecules.
In a more general case of realistic molecules, there should be cross-talk
between moments of different orders.  Although Grad has pointed out that the
coupling of this cross-talk is weak~\cite{Grad1958}, its role still needs to be
investigated in strong non-equilibrium situations.  Additionally, due to the
boundary-free nature of shock wave structures, the performance of SMRT with
gas-wall interaction remains to be studied.

\section*{Appendix}
\appendix
\section{Decomposition of a generic symmetric tensor}

\label{sec:decomp}

An arbitrary rank-$n$ symmetric tensor, $\bA{n}$, can be decomposed into its
irreducible components as~\cite{Spencer1970,Hannabuss1994}
\begin{equation}
	\label{eq:IrrDecomp}
	\bA{n} = \sum_{m=0}^{\lfloor n/2 \rfloor}\bA{n,m}\bdelta^m
\end{equation}
where $\bA{n,m}$ are trace-less symmetric tensors of rank $(n-2m)$ defined by
$\bA{n,m} = Q^{(n)}_m(\bA{n})$ where $Q^{(n)}_m$ are project operators given
by~\cite{Hannabuss1994}
\begin{equation}
	Q^{(n)}_m = B(n,m)Q^{(n-2m)}_0\tr^m,\quad
	Q^{(0)}_0 = 1,\quad\mbox{and}\quad
	Q^{(n)}_0 = \sum_{k=0}^{\lfloor n/2 \rfloor}D(n,k)\bdelta^k\tr^k
\end{equation}
where $\tr^n$ stands for applying the trace operator $n$ tines, and the constant
coefficients are
\begin{equation}
	\label{eq:bd}
	B(n,m) = \frac{n!(2n-4m+1)!!}{m!2^m(n-2m)!(2n-2m+1)!!}\quad\mbox{and}\quad
	D(n,m) = \frac{(-1)^mn!(2n-2m-1)!!}{m!2^m(n-2m)!(2n-1)!!}
\end{equation}
Shown in Table~\ref{tab:bd} are the numerical values of $B(n,m)$ and $D(n,m)$ at
the leading orders. With $D(n,m)$, the leading $Q^{(n)}_0$ can be easily
evaluated as
\begin{equation}
	Q^{(1)}_0 = 1,\quad
	Q^{(2)}_0 = 1 - \frac 13\bdelta\tr,\quad
	Q^{(3)}_0 = 1 - \frac 35\bdelta\tr,\quad
	Q^{(4)}_0 = 1 - \frac 67\bdelta\tr + \frac{3}{35}\bdelta^2\tr^2
\end{equation}
and further with $B(n,m)$, we have the irreducible tensors up to the fourth
order
\begin{subequations}
	\begin{align}
		\bA{2,0} & = B(2,0)Q_0^{(2)}\bA{2}      = \bA{2} - \frac 13\bdelta\tr\bA{2}                                   \\
		\bA{2,1} & = B(2,1)Q_0^{(0)}\tr\bA{2}   = \frac 13\tr\bA{2}                                                   \\
		\bA{3,0} & = B(3,0)Q_0^{(3)}\bA{3}      = \bA{3} - \frac 35\bdelta\tr\bA{3}                                   \\
		\bA{3,1} & = B(3,1)Q_0^{(1)}\tr\bA{3}   = \frac 35\tr\bA{3}                                                   \\
		\bA{4,0} & = B(4,0)Q_0^{(4)}\bA{4}      = \bA{4} - \frac 67\bdelta\tr\bA{4} + \frac 3{35}\bdelta^2\tr^2\bA{4} \\
		\bA{4,1} & = B(4,1)Q_0^{(2)}\tr\bA{4}   = \frac 67\tr\bA{4} - \frac 27\bdelta\tr^2\bA{4}                      \\
		\bA{4,2} & = B(4,2)Q_0^{(0)}\tr^2\bA{4} = \frac 15\tr^2\bA{4}
	\end{align}
\end{subequations}
It can be easily verified that Eq.~\eqref{eq:IrrDecomp} is satisfied.
\begin{table}[tb]
	\centering
	\caption{The leading coefficients $B(n,m)$, $D(n,m)$ as given by
		Eqs.~\eqref{eq:bd}}
	\begin{tabular}{l|ccc}
		\hline\hline
		      & $m=0$ &     $m=1$     &     $m=2$     \\ \hline
		$n=2$ & 1, 1  & $1/3$, $-1/3$ &               \\
		$n=3$ & 1, 1  & $3/5$, $-3/5$ &               \\
		$n=4$ & 1, 1  & $6/7$, $-6/7$ & $1/5$, $3/35$ \\ \hline\hline
	\end{tabular}
	\label{tab:bd}
\end{table}

\section{Convergence of Hermite Expansion for Maxwellian}

\label{sec:conver}

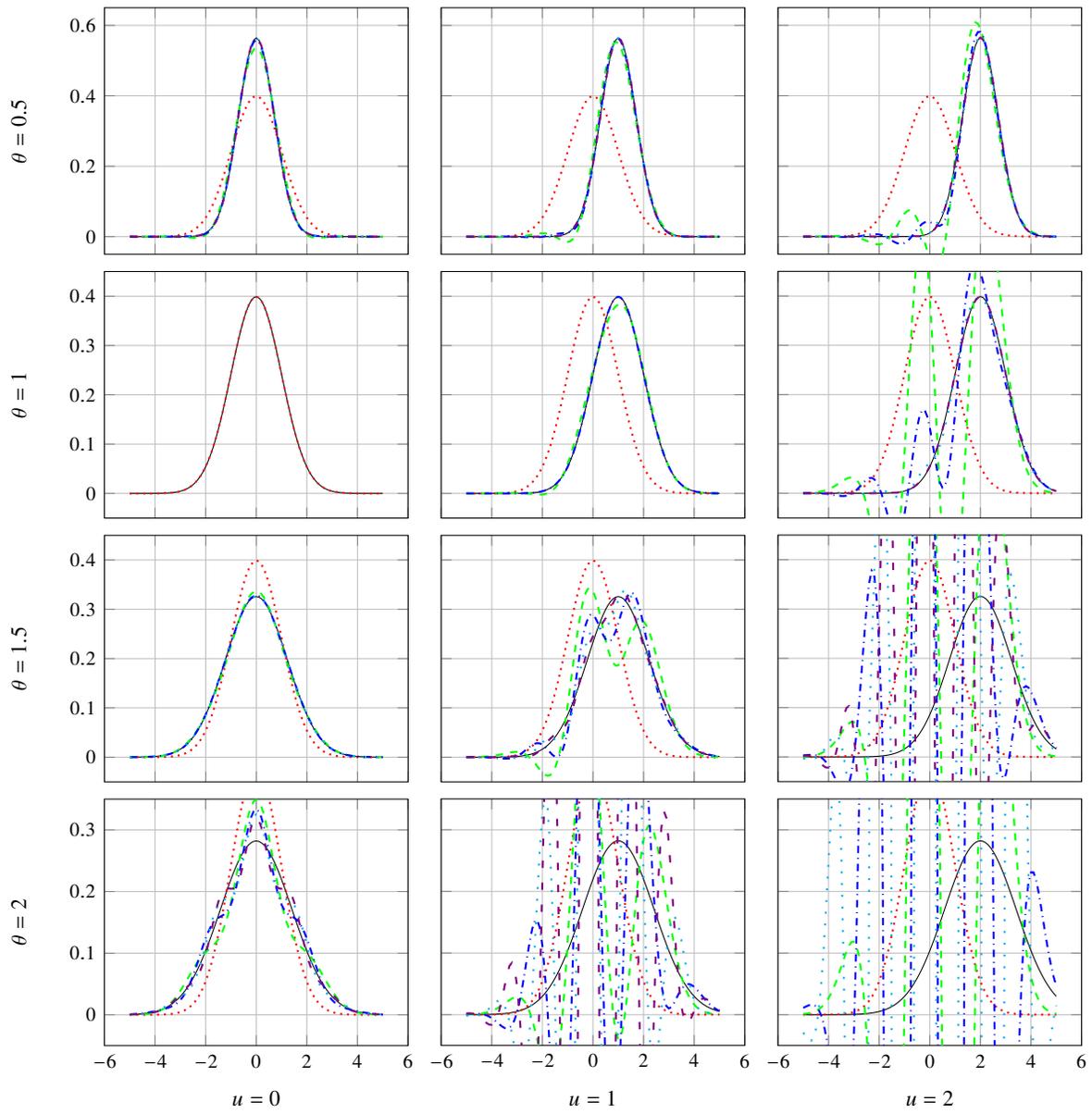
\begin{figure}
\centering
\begin{tikzpicture}
	\pgfplotsset{
		small,
		width=.36*\textwidth,
	    every axis plot post/.append style={mark=none},
	    every axis legend/.append style={font=\small},
	    cycle list={
	    	{solid},
	    	{red, dotted, thick},
	    	{green, dashed, thick},
	    	{blue, dashdotted, thick},
	    	{cyan, loosely dotted, thick},
	    	{violet, loosely dashed, thick}
	    },
	    ymin=-0.05,
	    ymax=0.65,
		grid=both
	}
	\matrix {
		\begin{axis}[xticklabel=\empty,ylabel={$\theta=0.5$}]
			\addplot table [x index=0, y index=1] {Fig/hermit_0.0_0.5.dat};
			\addplot table [x index=0, y index=2] {Fig/hermit_0.0_0.5.dat};
			\addplot table [x index=0, y index=3] {Fig/hermit_0.0_0.5.dat};
			\addplot table [x index=0, y index=4] {Fig/hermit_0.0_0.5.dat};
			\addplot table [x index=0, y index=5] {Fig/hermit_0.0_0.5.dat};
			\addplot table [x index=0, y index=6] {Fig/hermit_0.0_0.5.dat};
		\end{axis}
		&
		\begin{axis}[xticklabel=\empty,yticklabel=\empty]
			\addplot table [x index=0, y index=1] {Fig/hermit_1.0_0.5.dat};
			\addplot table [x index=0, y index=2] {Fig/hermit_1.0_0.5.dat};
			\addplot table [x index=0, y index=3] {Fig/hermit_1.0_0.5.dat};
			\addplot table [x index=0, y index=4] {Fig/hermit_1.0_0.5.dat};
			\addplot table [x index=0, y index=5] {Fig/hermit_1.0_0.5.dat};
			\addplot table [x index=0, y index=6] {Fig/hermit_1.0_0.5.dat};
		\end{axis}
		&
		\begin{axis}[xticklabel=\empty,yticklabel=\empty,legend pos=north west,legend cell align=left]
			\addplot table [x index=0, y index=1] {Fig/hermit_2.0_0.5.dat};
			\addplot table [x index=0, y index=2] {Fig/hermit_2.0_0.5.dat};
			\addplot table [x index=0, y index=3] {Fig/hermit_2.0_0.5.dat};
			\addplot table [x index=0, y index=4] {Fig/hermit_2.0_0.5.dat};
			\addplot table [x index=0, y index=5] {Fig/hermit_2.0_0.5.dat};
			\addplot table [x index=0, y index=6] {Fig/hermit_2.0_0.5.dat};
		\end{axis}
		\\
		\begin{axis}[xticklabel=\empty,ylabel={$\theta=1$},ymax=0.45]
			\addplot table [x index=0, y index=1] {Fig/hermit_0.0_1.0.dat};
			\addplot table [x index=0, y index=2] {Fig/hermit_0.0_1.0.dat};
		\end{axis}
		&
		\begin{axis}[xticklabel=\empty,yticklabel=\empty,ymax=0.45]
			\addplot table [x index=0, y index=1] {Fig/hermit_1.0_1.0.dat};
			\addplot table [x index=0, y index=2] {Fig/hermit_1.0_1.0.dat};
			\addplot table [x index=0, y index=3] {Fig/hermit_1.0_1.0.dat};
			\addplot table [x index=0, y index=4] {Fig/hermit_1.0_1.0.dat};
		\end{axis}
		&
		\begin{axis}[xticklabel=\empty,yticklabel=\empty,ymax=0.45]
			\addplot table [x index=0, y index=1] {Fig/hermit_2.0_1.0.dat};
			\addplot table [x index=0, y index=2] {Fig/hermit_2.0_1.0.dat};
			\addplot table [x index=0, y index=3] {Fig/hermit_2.0_1.0.dat};
			\addplot table [x index=0, y index=4] {Fig/hermit_2.0_1.0.dat};
			\addplot table [x index=0, y index=5] {Fig/hermit_2.0_1.0.dat};
			\addplot table [x index=0, y index=6] {Fig/hermit_2.0_1.0.dat};
		\end{axis}
		\\
		\begin{axis}[xticklabel=\empty,ylabel={$\theta=1.5$},ymax=0.45]
			\addplot table [x index=0, y index=1] {Fig/hermit_0.0_1.5.dat};
			\addplot table [x index=0, y index=2] {Fig/hermit_0.0_1.5.dat};
			\addplot table [x index=0, y index=3] {Fig/hermit_0.0_1.5.dat};
			\addplot table [x index=0, y index=4] {Fig/hermit_0.0_1.5.dat};
			\addplot table [x index=0, y index=5] {Fig/hermit_0.0_1.5.dat};
		\end{axis}
		&
		\begin{axis}[xticklabel=\empty,yticklabel=\empty,ymax=0.45]
			\addplot table [x index=0, y index=1] {Fig/hermit_1.0_1.5.dat};
			\addplot table [x index=0, y index=2] {Fig/hermit_1.0_1.5.dat};
			\addplot table [x index=0, y index=3] {Fig/hermit_1.0_1.5.dat};
			\addplot table [x index=0, y index=4] {Fig/hermit_1.0_1.5.dat};
			\addplot table [x index=0, y index=5] {Fig/hermit_1.0_1.5.dat};
			\addplot table [x index=0, y index=6] {Fig/hermit_1.0_1.5.dat};
		\end{axis}
		&
		\begin{axis}[xticklabel=\empty,yticklabel=\empty,ymax=0.45]
			\addplot table [x index=0, y index=1] {Fig/hermit_2.0_1.5.dat};
			\addplot table [x index=0, y index=2] {Fig/hermit_2.0_1.5.dat};
			\addplot table [x index=0, y index=3] {Fig/hermit_2.0_1.5.dat};
			\addplot table [x index=0, y index=4] {Fig/hermit_2.0_1.5.dat};
			\addplot table [x index=0, y index=5] {Fig/hermit_2.0_1.5.dat};
			\addplot table [x index=0, y index=6] {Fig/hermit_2.0_1.5.dat};
		\end{axis}
		\\
		\begin{axis}[ylabel={$\theta=2$},xlabel={$u=0$},ymax=0.35]
			\addplot table [x index=0, y index=1] {Fig/hermit_0.0_2.0.dat};
			\addplot table [x index=0, y index=2] {Fig/hermit_0.0_2.0.dat};
			\addplot table [x index=0, y index=3] {Fig/hermit_0.0_2.0.dat};
			\addplot table [x index=0, y index=4] {Fig/hermit_0.0_2.0.dat};
			\addplot table [x index=0, y index=5] {Fig/hermit_0.0_2.0.dat};
			\addplot table [x index=0, y index=6] {Fig/hermit_0.0_2.0.dat};
		\end{axis}
		&
		\begin{axis}[yticklabel=\empty,xlabel={$u=1$},ymax=0.35]
			\addplot table [x index=0, y index=1] {Fig/hermit_1.0_2.0.dat};
			\addplot table [x index=0, y index=2] {Fig/hermit_1.0_2.0.dat};
			\addplot table [x index=0, y index=3] {Fig/hermit_1.0_2.0.dat};
			\addplot table [x index=0, y index=4] {Fig/hermit_1.0_2.0.dat};
			\addplot table [x index=0, y index=5] {Fig/hermit_1.0_2.0.dat};
			\addplot table [x index=0, y index=6] {Fig/hermit_1.0_2.0.dat};
		\end{axis}
		&
		\begin{axis}[yticklabel=\empty,xlabel={$u=2$},ymax=0.35]
			\addplot table [x index=0, y index=1] {Fig/hermit_2.0_2.0.dat};
			\addplot table [x index=0, y index=2] {Fig/hermit_2.0_2.0.dat};
			\addplot table [x index=0, y index=3] {Fig/hermit_2.0_2.0.dat};
			\addplot table [x index=0, y index=4] {Fig/hermit_2.0_2.0.dat};
			\addplot table [x index=0, y index=5] {Fig/hermit_2.0_2.0.dat};
		\end{axis}
		\\
	};
\end{tikzpicture}
\caption{Convergence of Hermite expansions of the Maxwellian with different
	parameters.}
\label{fig:convergence}
\end{figure}

To illustrate the convergence speed of the Hermite expansion of a distribution
approximated by the Maxwellian, we display in Fig.~\ref{fig:convergence} the
Hermite expansions of the following 1D Maxwellian at various orders
\begin{equation}
	\label{eq:fexp}
	f \equiv\frac 1{\sqrt{2\pi\theta}}\exp\left[-\frac{(\xi-u)^2}{2\theta}\right]
	\cong \omega(\xi)\sum_{n=0}^N\frac{1}{n!}a^{(n)}\bH{n}(\xi)
\end{equation}
where $\bH{n}$ and $a^{(n)}$ have the closed forms of~\cite{Shan2021}
\begin{equation}
	\bH{n}(\xi) = \sum_{k=0}^{\lfloor n/2\rfloor}
	\frac{(-1)^kn!\xi^{n-2k}}{2^kk!(n-2k)!},
	\quad\mbox{and}\quad
	a{(n)} = \sum_{k=0}^{\lfloor n/2\rfloor}
	\frac{n!\left(\theta-1\right)^ku^{n-2k}}{2^kk!(n-2k)!}
\end{equation}
To be seen is that convergence speed strongly depends on the parameters. At $u =
0$ and $\theta = 1$, the weight function is the Maxwellian itself and the
expansion is exact.  The convergence speed deteriorates as $u$ and $\theta$ move
away from these values.  At $\theta = 2$, the theoretical convergence boundary,
the expansion fails to converge even for $u=0$.  The deterioration appears to be
slower for $\theta < 1$ than $\theta > 1$.  To give a rough hint about the
convergence speed of Eq.~\eqref{eq:fexp}, we plot in Fig.~\ref{fig:accuracy
	boundary} the smallest $N$ such that
\begin{equation}
	\int\left|f - \omega(\xi)\sum_{n=0}^N\frac{1}{n!}a^{(n)}\bH{n}(\xi)
	\right|d\xi < 0.01
\end{equation}

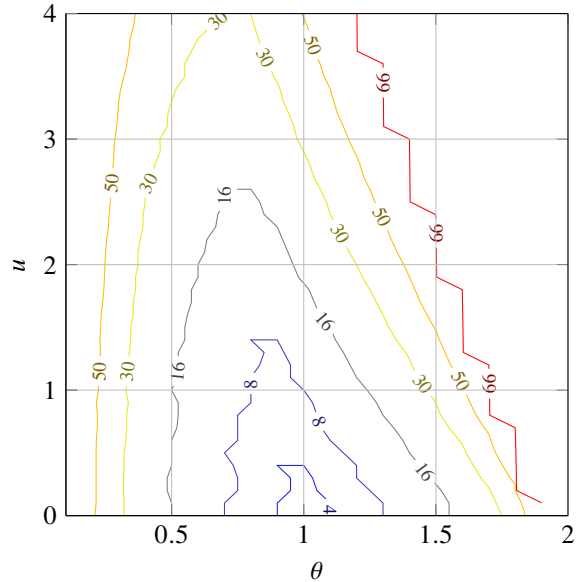
\begin{figure}[htbp]
	\centering
	\begin{tikzpicture}
		\begin{axis}[
			view={0}{90},
			xmin = 0.1, xmax = 2,
			ymin = 0, ymax = 4,
			grid = both,
			xlabel=$\theta$,
			ylabel=$u$,
			width=0.5*\textwidth,
			height=0.5*\textwidth
			]
			\addplot3 [contour gnuplot={levels={0, 4, 8, 16,30,50,99}}]
			table {Fig/Convergence_Hermite/boundary/convergence_contour.dat};			
		\end{axis}
	\end{tikzpicture}
	\caption{Convergence boundaries of absolute Hermite expansion for 1-D Maxwellian.}
	\label{fig:accuracy boundary}
\end{figure}

\section*{Funding Sources}

This work was supported by the National Natural Science Foundation of China
Grants No.~92152107 and No.~91952302, and by the Shenzhen Science and Technology
Program Grant No. KQTD20180411143441009.

\section*{Acknowledgments}

The authors thank Prof. L.~Wu for providing the Fast Spectral Method code for
benchmarking, and for many intriguing discussions.  S.Y. thanks Dr.~Y.~Shi for
helpful discussions.

\bibliography{sample}

\end{document}